\documentclass[10pt,letterpaper]{article}

\usepackage{fullpage}
\usepackage{cite}
\usepackage{algorithmic}
\usepackage{textcomp}

\usepackage{amsmath, graphicx, epsfig, color, amsfonts, algorithm, amssymb, mathtools, array, url, multirow, booktabs}
\usepackage{soul}
\usepackage{amsthm}
\usepackage{lineno}

\usepackage[font=footnotesize]{caption}
\usepackage{wrapfig}

\usepackage{subcaption}
\usepackage{float}

\usepackage[normalem]{ulem}

\usepackage[inline]{enumitem}

\usepackage{version,xspace}
\usepackage{tikz}
\usetikzlibrary{arrows.meta, positioning, fit, backgrounds, shapes.geometric}

\def\expt{\mathbb{E}}
\def\real{\mathbb{R}}

\def\naturals{\mathbb{N}}

\newcommand{\subscr}[2]{#1_{\textup{#2}}}
\newcommand{\supscr}[2]{#1^{\textup{#2}}}

\newcommand{\subj}{\text{subj. to}}

\newcommand{\norm}[1]{\left\lVert#1\right\rVert}

\def \bs {\boldsymbol}

\def\tran{^\top}

\def \mb {\mathbb}

\def\expt{\mathbb{E}}
\def\real{\mathbb{R}}

\def\naturals{\mathbb{N}}

\newcommand{\rev}[1]{\textcolor{black}{#1}}

\begin{document}
\title{Automated Curriculum Design for High-dimensional Human Motor Learning}
\author{
Ankur Kamboj$^1$, Rajiv Ranganathan$^2$, Xiaobo Tan$^1$, and Vaibhav Srivastava$^1$
\thanks{This work was supported by NSF Grant CMMI-1940950.\\
$^{1}$A. Kamboj ({\tt\small ankurank@msu.edu}), X. Tan ({\tt\small xbtan@msu.edu}), and V. Srivastava ({\tt\small vaibhav@msu.edu}) are with the Department of Electrical and Computer Engineering, Michigan State University, USA.\\
$^2$R. Ranganathan ({\tt\small rrangana@msu.edu}) is with the Department of Kinesiology and the Department of Mechanical Engineering, Michigan State University, USA.}
}
\date{}
\maketitle

\begin{abstract}
Designing effective practice schedules for high-dimensional motor learning tasks remains a challenge, especially when skill states are unobservable and task performance may not reflect the true learning. We propose an automated curriculum design framework that combines a human motor learning model and personalized real-time skill estimation with Stochastic Nonlinear Model Predictive Control in \emph{de-novo} (novel) motor learning paradigms. We validated our framework both through simulations and human-subject studies (N = 36) using a hand exoskeleton.  Our proposed approach accelerates skill acquisition by $\sim23\%$, and ${\sim17\%}$ when compared to a random curriculum and a performance heuristics-based curriculum, respectively. These significant gains in learning efficiency highlight the potential of model-based, individualized curricula for motor rehabilitation and complex skill training.
\end{abstract}

\section{Introduction} \label{sec:introduction}
The integration of robotics into physical rehabilitation has underscored the need for automated and adaptive training strategies that reduce reliance on human supervision and ensure consistent training and movement repeatability across rehabilitation sessions. Across fields such as education, neuroscience, and game design, task scheduling, also referred to as curriculum design, has been shown to significantly influence human motor skill acquisition and performance by optimizing the sequence of prescribed training tasks \cite{schmidt2018motor}. Several theories, such as the flow channel, challenge point, and difficulty adjustment, have informed curriculum design \cite{wadden2018individualized, zohaib2018dynamic, zhou2016learning}. However, in rehabilitation settings, especially those involving complex, high-dimensional motor systems like the human hand, curriculum design is still largely heuristic, relying on expert intuition rather than systematic, data-driven methods. This reliance limits scalability and effectiveness, particularly for non-expert users and in scenarios where motor redundancy makes it difficult to distinguish between efficient and effective strategies.
 
Estimating a learner’s underlying skill state during training is a non-trivial task. Common approaches rely on performance-based methods such as specific performance metrics~\cite{sungeelee2024interactive}, unsupervised feature extraction~\cite{lu2021evaluating}, or pre-study robotic assessments~\cite{metzger2014assessment}. \cite{ghonasgi2021capturing} leveraged various performance features for skill state evaluation in a human-skill curriculum Markov Decision Process. While these methods can track task outcomes, they often overlook latent aspects of motor learning in redundant high-dimensional systems. In such settings, improved short-term performance may not always correspond to skill learning~\cite{kantak2012learning}, especially because redundancy allows multiple movement strategies to achieve similar outcomes.

The predominant random task scheduling for high-dimensional de-novo motor learning tasks has been shown to result in sub-optimal learning for complex motor skills~\cite{shea1979contextual, ammar2023myth, frank2004contextual}.
Performance-based curriculum design has also been adopted, leading to improvements in motor learning efficiency~\cite{wadden2018individualized, choi2008performance, wang2022control}. However, current techniques either rely on expert-devised heuristics or follow task orders designed by an expert~\cite{basalp2019rowing, swanson2023optimized}. For high-dimensional motor tasks, redundancy makes distinguishing between performance and skill difficult, even for experts, limiting the generalizability of such approaches.
At worst, a poor curriculum design can severely hinder skill acquisition, underscoring the need for an adaptive and skill-informed curriculum strategy.
To overcome these limitations, we propose an SNMPC-based curriculum design framework that employs a particle filter to track a learner's real-time skill states informed by a motor learning model developed in our prior work~\cite{kamboj2024human}.
By characterizing high-dimensional skill evolution in real-time and its effect on task performance, our framework optimizes training task sequences over a lookahead horizon to minimize a combined skill-performance cost. This tailors training to the individual,  accelerating skill acquisition and improving performance.


%
%
\begin{figure*}[t!]
    \centering
    \begin{subfigure}{0.3\linewidth}
    \caption{}
	    \centering
        \includegraphics[width=1\linewidth, height=1.1\linewidth, keepaspectratio]{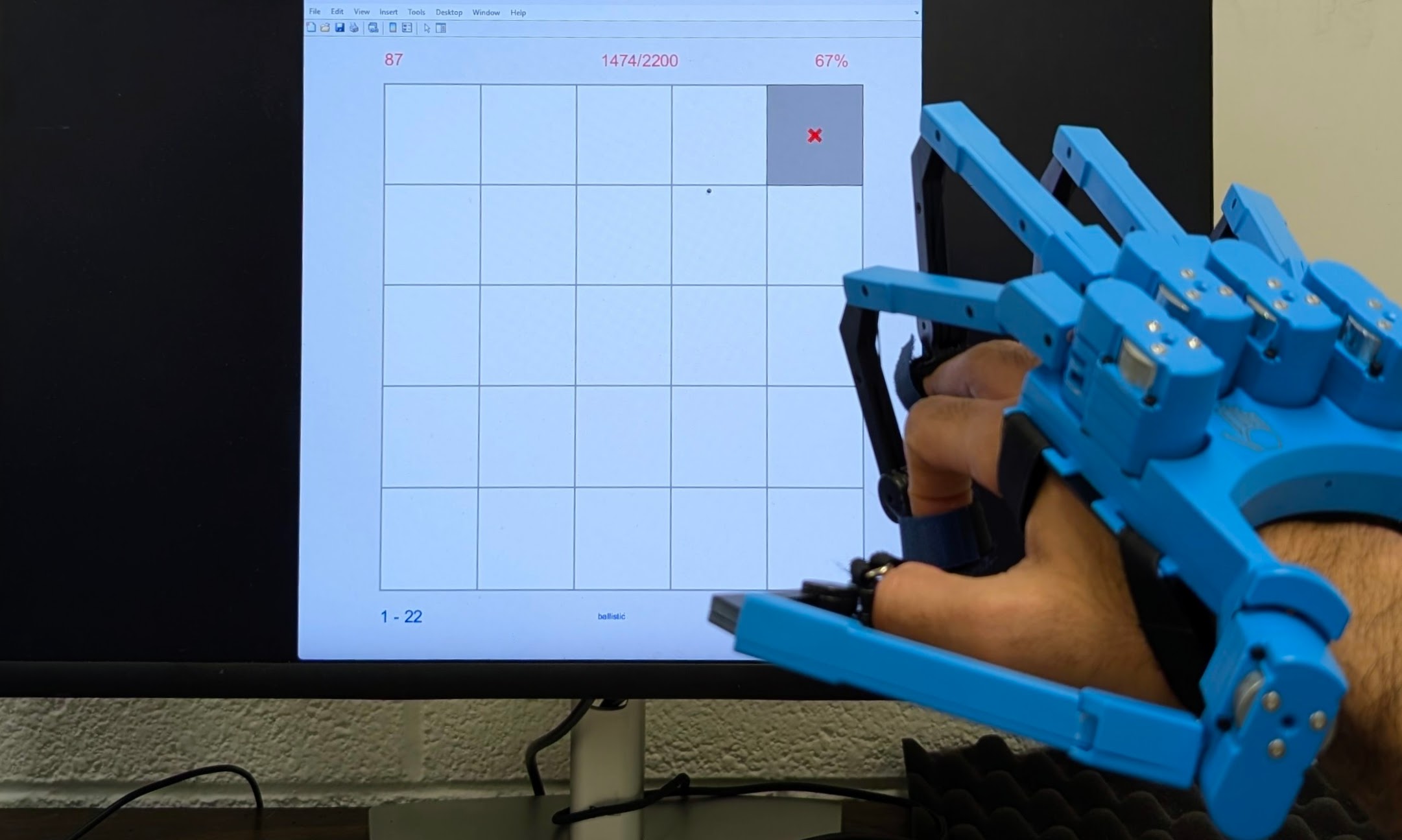}
        \label{fig:TCG_a} \vspace{0.15em}
    \end{subfigure}%
    ~
    \begin{subfigure}{0.3\linewidth}
    \caption{}
	    \centering
        \includegraphics[width=1\linewidth, height=1.1\linewidth, keepaspectratio]{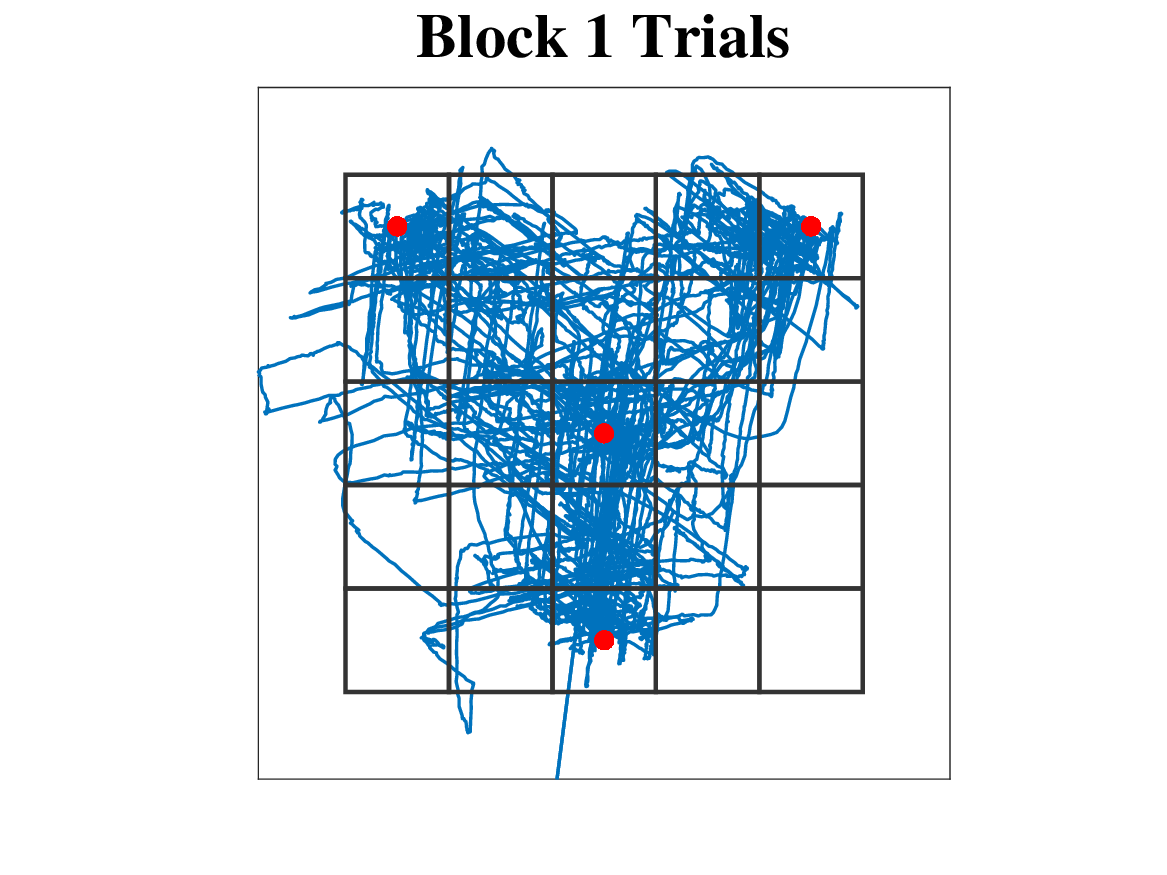}
    \end{subfigure}
    ~
    \begin{subfigure}{0.3\linewidth}
    \caption{}
	    \centering
        \includegraphics[width=1\linewidth, height=1.1\linewidth, keepaspectratio]{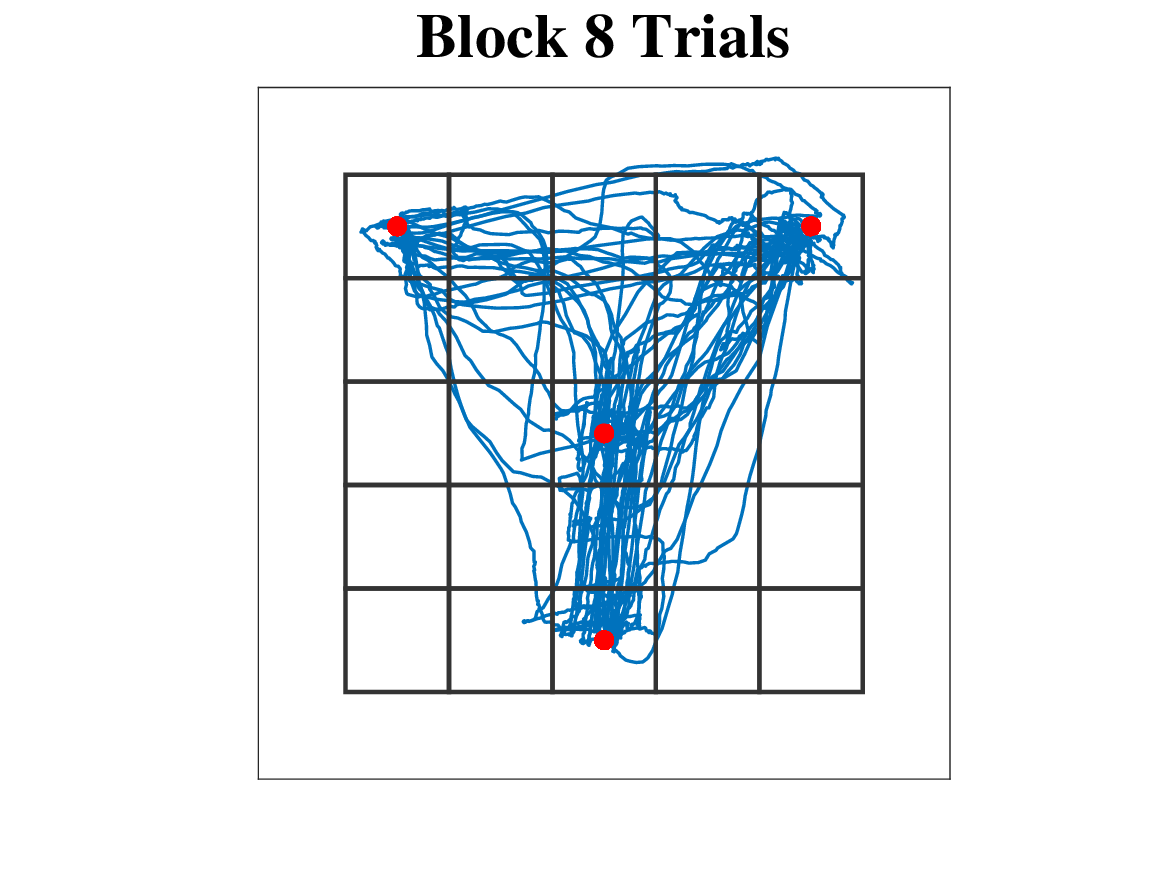}
    \end{subfigure}
    \vspace{-0.12in}
    \caption{\textbf{Target Capture Game:} (a) A participant playing the target capture game with the hand exoskeleton (SenseGlove DK1) strapped on the right hand. (b) The cursor trajectories during Block $1$, and (c) Block $8$ of the participant's target capture gameplay are shown. The red dots are the target points, and the $5\times 5$ unit grid is the task space shown to the participants. The grid units are computed to ensure participants' reachability to all grid squares.}
    \label{fig:TCG}
    \vspace{-0.25in}
\end{figure*}
Building on the preliminary findings of our SNMPC-based curriculum design~\cite{kamboj2025expediting}, this work investigates the potential of the skill-informed curriculum design in accelerating skill acquisition for high-dimensional, de-novo motor tasks. We distinguish this study from previous work through the following advancements:
\begin{enumerate*}[label=(\roman*)]
    \item we demonstrate the strength of our proposed approach without the need for prior model-fitting sessions, thus removing any skill carry-over effects from these sessions and ensuring a true baseline at the start of learning,
    \item we expand the participant pool and introduce a heuristics-based group to benchmark performance against random and SNMPC-based curricula,
    \item the SNMPC cost function now includes performance metrics for increased flexibility in determining the performance-skill trade-off,
    \item we employ Uncontrolled Manifold analysis to evaluate how effectively participants utilize learned forward mappings by analyzing movement variability in the task space, and
    \item we study the system's robustness to model uncertainties and delineate the choice of a nonlinear filter for skill state estimation.
\end{enumerate*}
In summary, our contributions are threefold: 
\begin{enumerate}
\item We design a generalizable model-based online skill estimation framework using a human motor learning (HML) model and a particle filter.
\item We propose an SNMPC framework that leverages the real-time skill state estimate to design an optimal curriculum based on skill evolution and its effect on task performance.
\item Through simulations and human studies, we demonstrate the efficacy of SNMPC curriculum design over popular approaches (random curriculum and performance heuristics-based curriculum), in expediting motor skill learning and enhancing task performance outcomes.
\end{enumerate}

The remainder of the paper is organized as follows. The motor learning experimental setup is presented in Section \ref{sec:exp_setup}. Section \ref{sec:curriculum_algo} details the computational model of human motor learning leveraged in this paper, followed by the proposed model-based skill state estimation framework and curriculum design algorithms. The simulation studies and results from human experiments are presented in Section \ref{sec:results}. Some discussions on the results, techniques adopted, and potential future directions are presented in Section \ref{sec:discussion}, followed by Methods in Section \ref{sec:materialandmethods}.

\section{Motor Learning Experiment Setup}  \label{sec:exp_setup}
Our experiment involves healthy participants playing a target capture game \cite{kamboj2024human} using a non-invasive hand exoskeleton (SenseGlove DK1, manufactured by SenseGlove, The Netherlands) that records $20$ finger joint angles. Let $\bs q \in \real^{20}$ be the vector of these finger joint angles, with the velocities $\bs u = \dot{\bs q}$. These joint velocities are mapped to $2$D cursor velocities $\dot{\bs x} \in \real^2$ using a body-machine interface via the mapping matrix $C$ as
\begin{align}\label{base_dyn}
\dot{\bs x} = C \bs u. \quad
\end{align}
Participants are tasked with playing the target capture game, which involves capturing targets on a computer screen (Fig. \ref{fig:TCG_a}) by moving their $20$ finger joints to guide the $2$D cursor towards these targets. 
This task involves motor learning in a high-dimensional system ($20$-D finger joint space) guided by low-dimensional feedback ($2$-D screen space), effectively training their nervous system to master the underlying mapping matrix $C$.
Experiment sessions are divided into three phases: calibration, familiarization, and training.

During the \emph{calibration phase}, the participants are asked to form the American Sign Language hand postures, which are used to design the personalized mapping matrix $C$. To maintain task difficulty, the second and third principal components are extracted via Principal Component Analysis on the centered calibration data and scaled by the square root of their eigenvalues to form the two rows of the mapping matrix $C$. The task space is represented as a $5\times 5$ unit game window, centered on the participant's mean posture. One unit is the width of a grid square, and is computed such that the game window accommodates a $1$-standard deviation of the calibration data points, ensuring the whole task space is reachable without extreme motions.

In the \emph{familiarization phase}, participants are asked to move the cursor freely for $\sim6~$s to acclimate to the interface without inducing motor learning, and also to ensure that the entire gameplay grid is reachable without resorting to extreme hand postures.

The \emph{training phase} is composed of $8$ blocks with $60$ trials each of the target capture gameplay. Each trial involves a reaching movement of screen cursor to one of four targets squares with centers located at $(0.5, 4.5), (2.5, 0.5), (2.5, 2.5), \textup{ and } (4.5, 4.5)$ units, as shown in Fig. \ref{fig:TCG}\footnote{A video demonstrating a participant performing this experiment can be found at \url{https://youtu.be/WIDRwqpE9Sk}.}, where $(0,0)$ is the bottom left corner of the game window. A target is considered captured if the cursor remains stable (variance $<0.0025$ units) within the target square for $15$ consecutive samples (at $100$ Hz), and a new target appears after the capture. To motivate participants, a scoring system based on accuracy, trajectory straightness, and time is displayed. If a capture exceeds $2~$s, the target square highlights red and a time penalty is incurred. During a trial, the cursor is allowed to go outside the game window bounds, but participants see it only at the game window edges, with no feedback on how far out of the game window the cursor is. It is to be noted that since game window units are calculated from the scaling data during the calibration phase, they are different for different participants.

As participants practiced, they gradually learned the inverse mapping required to control the cursor via coordinated finger motions. Due to the many-to-one structure of $C$, multiple synergistic finger movements yield the same cursor motion. This high redundancy makes the motor learning aspect of the task especially difficult.
\section{Skill State Estimation and Curriculum Design}  \label{sec:curriculum_algo}
Curriculum design for high-dimensional, novel motor tasks is hindered by two factors: the inability to directly measure latent skill state, which often leads to a reliance on sub-optimal, low-dimensional performance metrics, and the difficulty of providing effective visual guidance in high degrees of freedom (DoF) spaces. To address these, we propose a data-driven framework that estimates latent skill states online to prescribe personalized curriculum design, posed as the selection of target points in our target capture game. By leveraging a computational motor learning model to infer skill evolution, the framework prescribes subject-specific curricula that enhance both learning and task performance.


\subsection{Human Motor Learning Computational Model}
Building on our previous work~\cite{kamboj2024human}, we use a computational model of human motor learning (HML) that captures the coordinated behavior of high-dimensional motor control using low-dimensional \emph{motor synergies} (coordinated joint movements) in healthy human participants.
These synergies are extracted from each participant's calibration data using principal component analysis, yielding four leading principal modes that define the mapping matrix $C=W\Phi$, where rows of $\Phi$ represent synergies, and $W$ denotes their weights.

Drawing from the internal model theory of motor learning~\cite{jordan1992forward, wolpert1995internal} and adaptive control, the HML model consists of fast and slow-adapting forward and inverse learning models.
The HML model is posed as the nonlinear stochastic dynamics
\begin{align} \label{HML_model}
    \dot{\bs \zeta}(t) = f\left(\bs \zeta(t), \bs \upsilon, \bs \omega(t)\right),
\end{align}
where $\bs \zeta$ is the concatenated model state $\bs \zeta(t) = \begin{bmatrix}
    {\bs x}\tran(t),
    {\hat{W}}\tran(t),
    {\delta \bs q}\tran(t),
    {\bs u}\tran(t),
    {\bs q}\tran(t)
\end{bmatrix}\tran$, $\hat{W}(t)$ is the participant's implicit estimate of synergy weights.
$\bs x$ is the cursor position, $\delta \bs q$ represents filtered joint angle increments, $\bs \omega(t)$ represents the process noise, and $\bs v = \supscr{\bs x}{des}(t)$ is the desired (target point) cursor position that the participant seeks at time $t$. 

Further details on the full HML model (functional form of $f(\cdot)$), synergy extraction, and parameter estimation are available in the Appendix. This fitted HML model serves as the core of our skill estimation framework, enabling real-time inference of latent skill states that are subsequently used for curriculum optimization to assist participants.

\subsection{Estimating Human Skill State}
Under our experimental setup, participants are required to learn coordinated finger joint motions that are consistent with their mapping matrix $C$ to control the cursor to capture the various targets. Consequently, the participant's implicit estimate of the mapping matrix $\hat{C}(t) = \hat{W}(t)\Phi$ is treated as the proxy for human skill state at time $t$. Since $\Phi$ is fixed per participant, tracking $\hat{W}(t)$ suffices as a proxy of learning progression\footnote{Although we use the concept of motor synergies, the idea is not central to the proposed skill state estimation and curriculum design. $\hat{C}$ could be used in place of $\hat{W}$, albeit with an increased computational overhead.}. The system evolution dynamics during game trial $j$ are captured by the nonlinear stochastic dynamics model \eqref{HML_model}. Since cursor position $\bs x(t)$ and finger joint angles $\bs q(t)$ are available online as observations during the experiment, we write the observation model as
\begin{align}   \label{eq:PF_obs}
    \bs y(t) = \left[\bs x\tran(t), \bs q\tran(t) \right]\tran = g\left(\bs \zeta(t), \bs \nu(t)\right),
\end{align}
where $\bs \nu(t)$ is the measurement noise.
Using \eqref{HML_model} as the state dynamics with the fitted model parameters and \eqref{eq:PF_obs} as the observation model, we can design an individualized nonlinear filter~\cite[Chapter 3, 4]{thrun2005probabilistic} to estimate the participant's skill state $\hat{W}(t)$. We evaluated multiple nonlinear filtering techniques, including the extended Kalman filter, the unscented Kalman filter, and the particle filter for skill state estimation. Due to the nonlinear nature of evolution dynamics, the particle filter consistently outperformed the other methods in tracking accuracy and robustness (nonlinear filters have been compared in the Appendix).
We thus adopt the particle filter to estimate the latent skill state $\hat{W}$ from observed cursor and joint trajectories during training.

\subsection{Curriculum Design Algorithms}
Standard practice employs uniformly random target prescription, a non-adaptive strategy that often results in inefficient training where learners may master natural motions but struggle with others. While expert-guided heuristics offer an alternative, identifying reliable performance metrics for high-dimensional motor systems is challenging. In these redundant systems, immediate task success is often a poor proxy for genuine skill acquisition~\cite{schmidt2018motor}. This motivates an adaptive curriculum design that accounts for learning dynamics to improve task performance. We evaluate two curriculum design methods: a performance-based heuristic and a model-based SNMPC approach.

\subsubsection{Performance Heuristics-based Manual Target Sequencing Curriculum Design} \label{sec:manualTS}
Informed by pilot studies, our heuristic curriculum uses two key performance metrics: Reaching Error (\texttt{RE}) and Straightness of Trajectory (\texttt{SoT}). We define \texttt{RE} as the Euclidean distance of the cursor from the target point, measured at the end of the movement or $2~$s (prescribed trial duration) after the movement starts, whichever is sooner. \texttt{SoT} is the ratio of the cursor trajectory's maximum perpendicular deviation from the straight line path (connecting the start and end points) to the length of that path.

Target selection is guided by a simple optimization procedure aimed at minimizing a participant's performance cost. The performance cost associated with each \emph{target pair} $p$ (a target pair consists of the current and the next prescribed target) is computed from a running average of \texttt{RE} and \texttt{SoT} as
\begin{align}
    \textup{Cost}_p = \norm{\subscr{\texttt{RE}}{avg, p}} + 10 * \norm{\subscr{\texttt{SoT}}{avg, p}}.
\end{align}
We weigh \texttt{SoT} more heavily because metrics derived from movement smoothness are effective quantifiers of motor learning improvements in rehabilitative contexts \cite{lu2021evaluating}. At the end of each trial, the target with the highest performance cost is selected for the next trial, thereby focusing practice on more challenging movements.

\subsubsection{Stochastic Nonlinear Model Predictive Control for Target Sequencing Curriculum Design}
The heuristic sequencing strategy discussed above is greedy and myopic, and might not lead to the best long-term outcomes. The training is a sequential process wherein the choice of the target in each trial influences the future distributions of motor skill states. Therefore, an effective curriculum design algorithm should plan ahead and account for the uncertainty in the evolution of motor skills. To this end, we employ SNMPC for target sequencing.

Since the target point is fixed for each trial, we rewrite our system dynamics \eqref{HML_model} as \emph{trial-to-trial dynamics} to aid in curriculum algorithm design. We write the trial-to-trial evolution of \eqref{HML_model} as the stochastic discrete nonlinear system
\begin{align}
    {\bs \zeta}_{j+1} = \Psi(\bs \zeta_j, \bs \upsilon_{j+1}, \Omega_j), \qquad
    \bs y_j = \Theta(\bs \zeta_j, \rm N_j),
\end{align}
where, $j \in \naturals$ is the trial number, $\bs \zeta_j$ is the system state at the end of trial $j$, $\bs \upsilon_{j+1} \in \mb U \subset \real^{2}$ is the target point for trial $j+1$, $\mb U$ is a non-empty, finite and countable set of admissible target points, and $\bs y_j$ is the system output. The nonlinear functions $\Psi$ and $\Theta$ describe the trial-to-trial dynamics, where $\Omega_j$ and $\bs \rm N_j$ represent the stochastic process and measurement noise over trial $j$. $\Psi$ and $\Theta$ are not known in closed forms, but model \eqref{HML_model} can be simulated to generate samples from them.

We define $P \in \naturals$ as the prediction (lookahead) horizon and define the decision variable as a sequence of $P$ target points
\begin{align}
    \bs \pi := \{ \bs \upsilon_{j+1}, \bs \upsilon_{j+2}, \ldots, \bs\upsilon_{j+P}\},
\end{align}
where each $\bs \upsilon_i \in \mb U, i = \{j+1, \ldots, j+P\}$ is the candidate target point for trial $i$.
The performance after each trial is quantified by a cost function $\ell_j$, which incorporates the error in estimated synergy weights $\tilde{W}_j$, reaching error $\texttt{RE}_{\upsilon}$, and straightness of trajectory $\texttt{SoT}_{\upsilon}$ as
\begin{align}
    \ell_j = \beta_W \norm{\tilde{W}_j}_2 + \beta_{RE}\norm{\texttt{RE}_{\upsilon}}_2 + \beta_{SoT}\norm{\texttt{SoT}_{\upsilon}}_2,
\end{align}
where $\beta$'s are the weight coefficients.
Thus, the $P$-step lookahead value function as a measure of the participant's task learning and performance is given by 
\begin{align}   \label{Vp_Wtilde}
    V_P(\bs \zeta_j, \bs \pi) = \expt_{\bs \zeta_{j+1}, \ldots, \bs \zeta_{j+P}} 
    \left[ \sum_{i=j+1}^{j+P-1} \ell_i + \beta_P \ell_{j+P} \right],
\end{align}
where $\ell_i$ is the stage cost incurred after trial $i$, $\beta_P$ is the weight on terminal cost, and the expectation is calculated using Monte Carlo approximation, computed by simulating realizations of $\Psi$ and $\Theta$, over the $P$-step lookahead horizon. Given the current system states $\bs \zeta_{j}$, the SNMPC algorithm solves the following stochastic optimization problem
\begin{equation}    \label{SNMPC}
    \begin{aligned}
        & \min \limits_{\bs \pi}\  V_P(\bs \zeta_{{j}}, \bs \pi), \\
        \subj\quad &\bs \zeta_{{i+1}} = \Psi(\bs \zeta_{i}, \bs \upsilon_{i+1}, \Omega_i),   \\
        &\bs \upsilon_i \in \mb U, i = \{j, \ldots, j+P-1\}.
    \end{aligned}
\end{equation}
Let $\bs \pi^*$ be the optimal policy obtained by solving \eqref{SNMPC}, and $V_P^*(\bs \zeta_j) = V_P(\bs \zeta_j, \bs \pi^*)$ denote the optimal value function.
At the end of trial $j$, the $P$-step lookahead horizon SNMPC selects the first target point from the optimal $P$-step target sequence $\bs \pi^*$ obtained by solving \eqref{SNMPC} as the target for the $(j+1)$th trial. We implement this optimization by performing a tree search over all admissible target sequences $\bs \pi$.

Owing to the structure of the mapping matrix $C$ and the chosen configuration of target points, the SNMPC algorithm \eqref{SNMPC} sometimes drives the optimizer to alternate between just two targets. Participants reported during pilot testing that this repetitive sequence made the game monotonous and fatiguing. To counteract this, we define the state action value function
\begin{align}
    Q(\bs \zeta_j, \bs \upsilon_{j+1}) = V_1(\bs \zeta_j, \bs \upsilon_{j+1}) + \expt_{\bs \zeta_{j+1}} \left[V_{P-1}^*(\bs \zeta_{j+1})\right],
\end{align}
and use the $\text{softmin}_{\tau}$ operator to define the probability distribution over the feasible targets for the next trial,
\begin{align}   \label{eq:policy_prob}
    p_{\bs \upsilon_{j+1}}\left(Q(\bs \zeta_{j}, \bs \upsilon_{j+1})\right) \propto \text{softmin}_{\tau}\left(Q(\bs \zeta_{j}, \bs \upsilon_{j+1})\right),
\end{align}
thus adding stochasticity to the target selection. 
The $\text{softmin}_{\tau}$ operator is defined as
\begin{align}
    \text{softmin}_{\tau}(\bs z_i) &= \frac{\exp(-(\bs z_i-\subscr{\bs z}{min})/\tau)}{\sum_k \exp(-(\bs z_k-\subscr{\bs z}{min}/\tau)}, \notag \\
    \subscr{\bs z}{min} &= \min_k(\bs z_k), \notag
\end{align}
with $k$ indexing the feasible target points for the next trial.
As $\tau \rightarrow 0$, the policy becomes deterministic, that is, the policy obtained by solving \eqref{SNMPC}, whereas larger $\tau$ values inject more randomness.
In our experiments, we set $\tau = 0.2$, which effectively breaks up repetitive target sequence prescriptions while preserving the optimality benefits of SNMPC. Adding controlled randomness through $\tau$ also provides robustness to model inaccuracies, as shown in Section~\ref{sec:discussion:MPC_robustness}.

\begin{figure}[b!]
\vspace{-0.2in}
    \centering
    \begin{subfigure}{0.45\linewidth}
        \centering\caption{}
        \includegraphics[width=1\linewidth, height=1\linewidth, keepaspectratio]{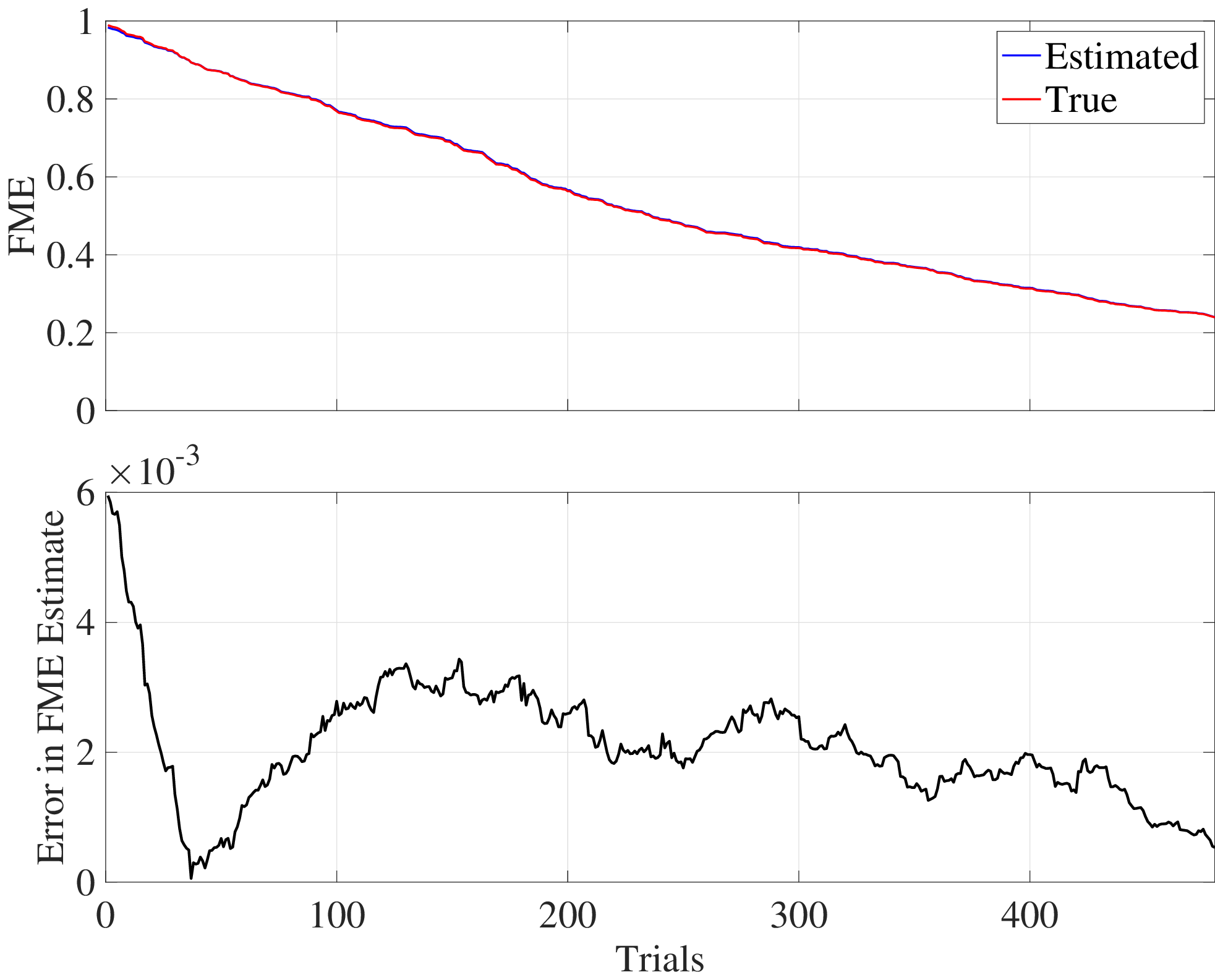}        
        \label{fig:pf_sim}
    \end{subfigure}\hspace{-0.3em}%
    ~
    \centering
    \begin{subfigure}{0.5\linewidth}
        \centering
        \caption{}
        \includegraphics[width=1\linewidth, height=1\linewidth, keepaspectratio]{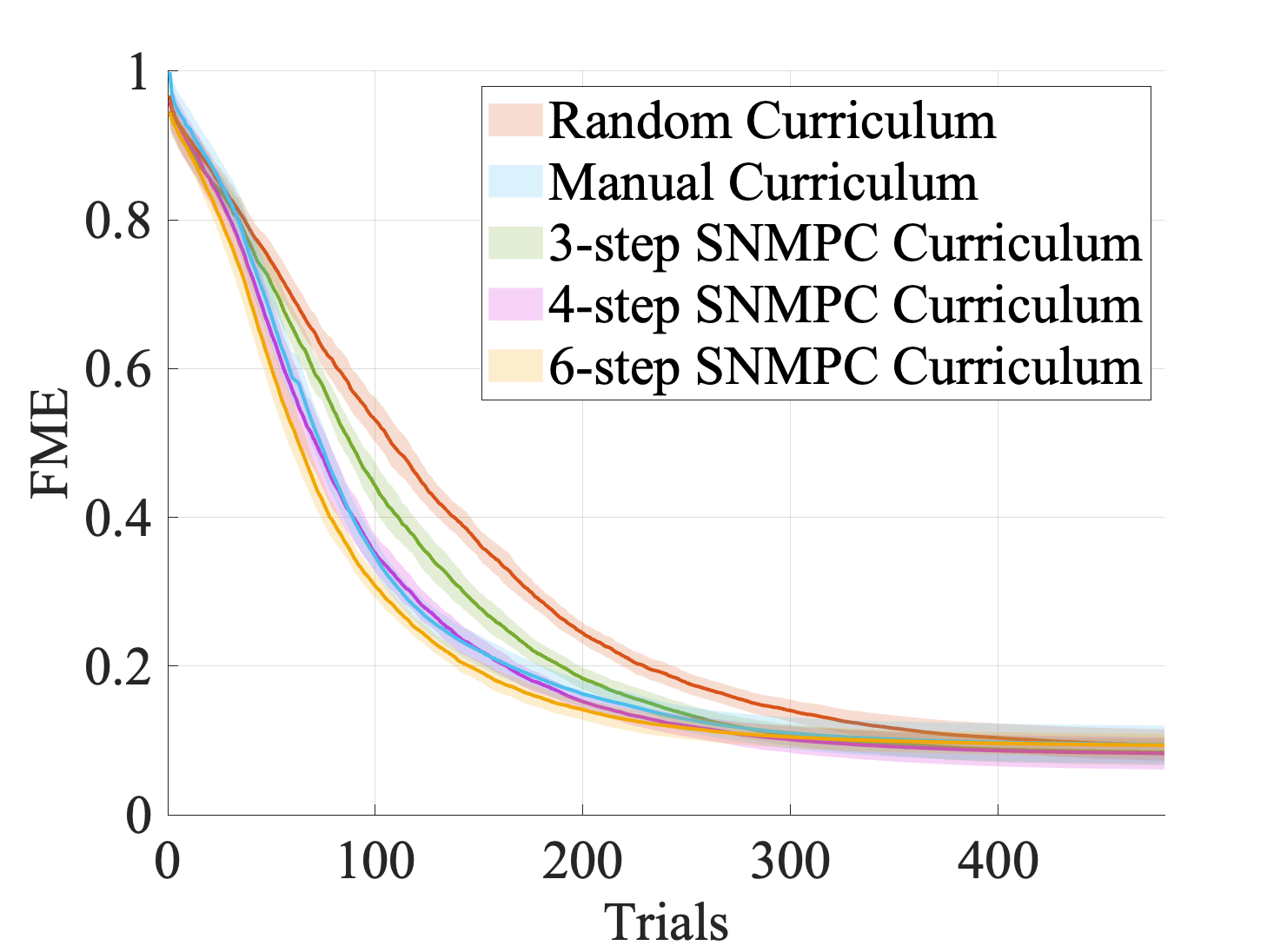}        
        \label{fig:FME_sim}
    \end{subfigure}\hspace{-1.2em}%
    \vspace{-0.18in}
    \caption{\textbf{FME in Simulation:} (a) FME evolution can be effectively captured by the particle filter using output data from a fitted HML model. (b) The SNMPC curriculum takes $\sim 80$ fewer trials compared to the random curriculum to achieve $80\%$ learning on the mean HML model, indicating expedited learning using our proposed curriculum design in simulation. Curves plotted are the mean with $95\%$ confidence interval over $10$ Monte Carlo runs.}
    \label{fig:FME_sim_results}
\end{figure}
\begin{figure*}[t!]
    \centering
    \begin{subfigure}{0.34\linewidth}
	    \centering
        \caption{}
        \includegraphics[width=1\linewidth, height=1\linewidth, keepaspectratio]{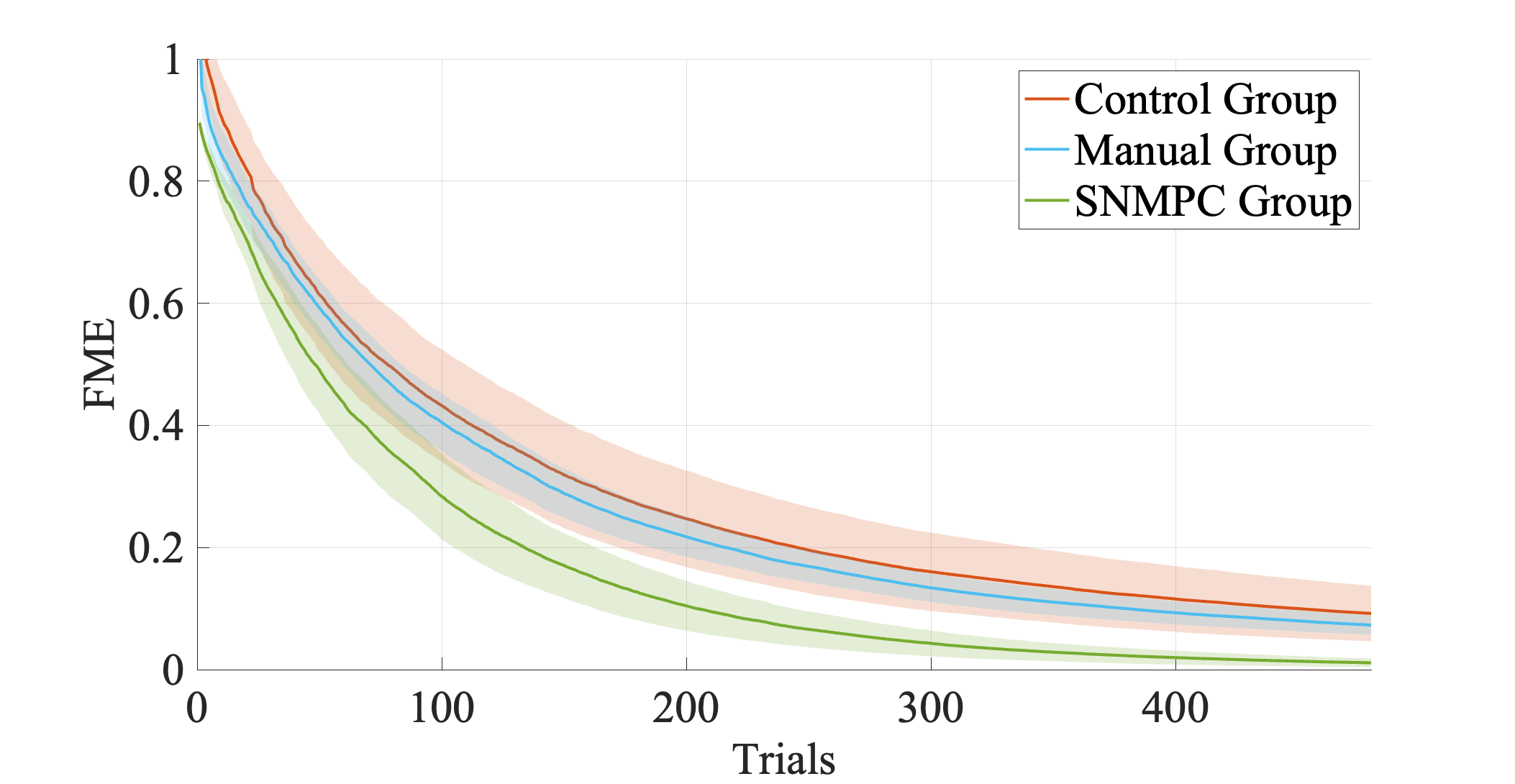}
        \label{fig:FME_exp}
    \end{subfigure}\hspace{-1.2em}%
    ~
    \centering
    \begin{subfigure}{0.34\linewidth}
	    \centering
        \caption{}
        \includegraphics[width=1\linewidth, height=1\linewidth, keepaspectratio]{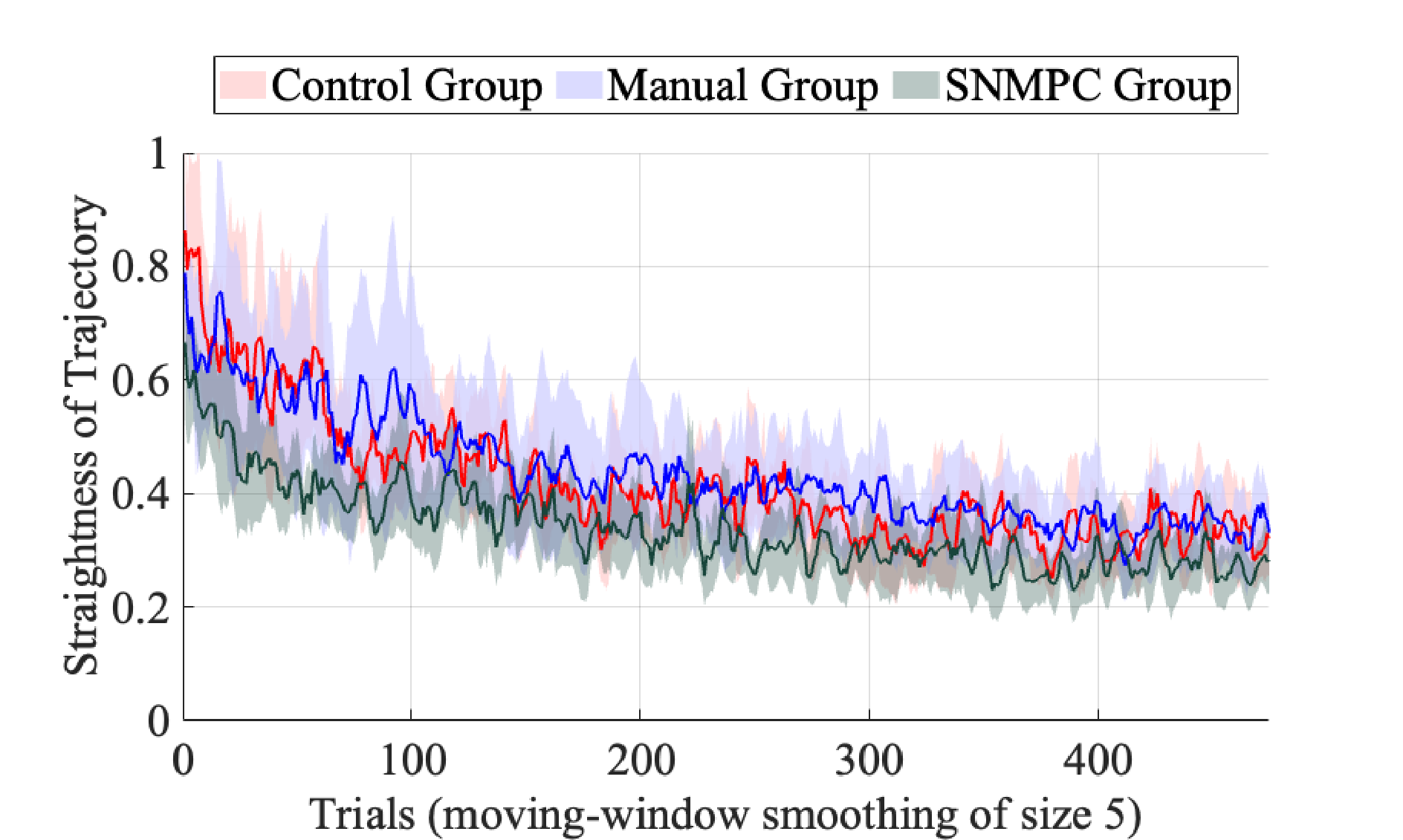}
        \label{fig:SOT_curves}
    \end{subfigure}\hspace{-1.2em}%
    ~
    \centering
    \begin{subfigure}{0.34\linewidth}
	    \centering
        \caption{}
        \includegraphics[width=1\linewidth, height=1\linewidth, keepaspectratio]{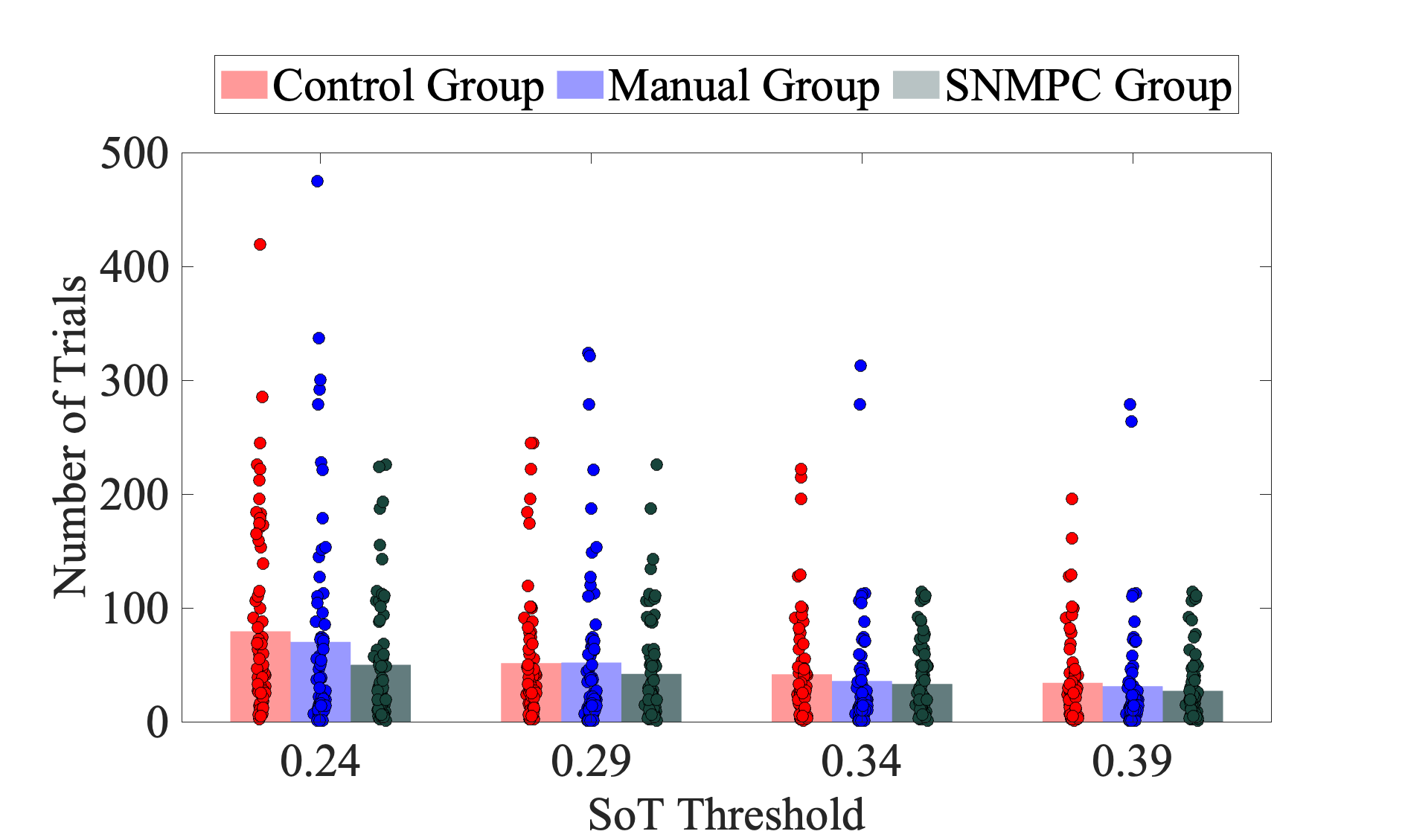}
        \label{fig:SOT_threshold}
    \end{subfigure}\hspace{-1.2em}%
    \vspace{-0.12in}
    \caption{\textbf{Task and Learning Performance:} (a) Learning error evolution for participants playing the game under the three curricula shows that on average, the SNMPC group takes $109$ and $80$ fewer trials than the control and manual groups, respectively, to achieve $80\%$ learning. (b) \texttt{SoT} curves of participants from the three groups, and (c) the mean number of trials (and the scatter plot across participants and target pairs) to converge to different \texttt{SoT} thresholds show that the SNMPC group converged the fastest, followed by the manual group. The p-values of two-tailed tests from Linear Mixed Model fits indicate a statistically significant decrease in \texttt{SoT} across trials and in the number of trials required by the SNMPC group for low \texttt{SoT} values. Each scatter plot point represents the number of trials taken by a participant to achieve the \texttt{SoT} threshold for a target pair.
    Curves plotted are the mean with $95\%$ confidence intervals.}
    \label{fig:exp_results}
    \vspace{-0.25in}
\end{figure*}

\section{Results}   \label{sec:results}
We evaluated the efficacy of our curriculum design framework through simulation studies and human-subject experiments.

%
%
%

\subsection{Simulation Results}
In simulation studies, we use the HML model fitted to one of the participants' data as a proxy for a human playing the game. The sequence of target pairs prescribed to the HML model modulates how the learner's estimated skill state, modeled by $\hat{W}$, evolves through and across the trials. This affects the learning performance, which is captured by the Forward Modeling Error (\texttt{FME}) metric. It is defined as
\begin{align}
    \texttt{FME} = \left. \norm{C - \hat{C}}_2 \middle/ \norm{C}_2\right.,
\end{align}
quantifying the learning error, that is, convergence of the human estimate of the forward mapping matrix $\hat{C}$ to the true mapping $C$. Evolution of \texttt{FME} thus captures how well a human is learning the novel motor task. 
Fig. \ref{fig:pf_sim} shows the efficacy of the designed particle filter in estimating the learning state $\hat{C}$ and capturing the learning error trend across trials in simulation.
The performance of three curricula was evaluated through simulation first, where a fitted HML model played the target capture game across $10$ Monte Carlo repetitions per group. These groups were defined by their scheduling logic: random (control group), heuristics (manual group), and SNMPC-based (SNMPC group).

Fig. \ref{fig:FME_sim} shows the learning error trend of the control, manual, and $3$, $4$, and $6$-step lookahead SNMPC groups in simulation. On average, both manual and $4$-step SNMPC groups take $\sim 60$ fewer trials ($\sim1$ block) than control group to achieve $80\%$ learning (\texttt{FME} $\approx0.2$). Moreover, \texttt{FME} decreases faster with increasing lookahead horizon. Thus, SNMPC can expedite motor learning for our novel learning task simulation. Given the diminishing improvements in learning and concurrent increase in computation with increasing SNMPC lookahead horizon, we used $4$-step SNMPC in our human-subject experiments.

\subsection{Experimental Results}
\subsubsection{Skill Learning Performance}
Fig. \ref{fig:FME_exp} illustrates the evolution of the learning error for the three participant groups. The average \texttt{FME} curves depict an appreciable trend towards expedited task learning when targets are sequenced according to the proposed SNMPC-based curriculum design framework.
Notably, on average, the SNMPC group requires $109$ and $80$ fewer trials compared to the random and manual groups, respectively, to achieve an \texttt{FME} of approximately $0.2$. This corresponds to a $\sim27\%$ and a $\sim17\%$ respective improvement to achieve $80\%$ learning. Furthermore, SNMPC curriculum participants clearly outperformed the other two groups throughout the motor task, achieving much lower learning error by the end. Recall that the particle filter yields the expected $\hat{W}$ as the estimate, leading to smooth curves in Fig. \ref{fig:FME_exp}.
\begin{figure*}[t!]
    \centering
    \begin{subfigure}{0.34\linewidth}
	    \centering
        \caption{}
        \includegraphics[width=1\linewidth, height=1.1\linewidth, keepaspectratio]{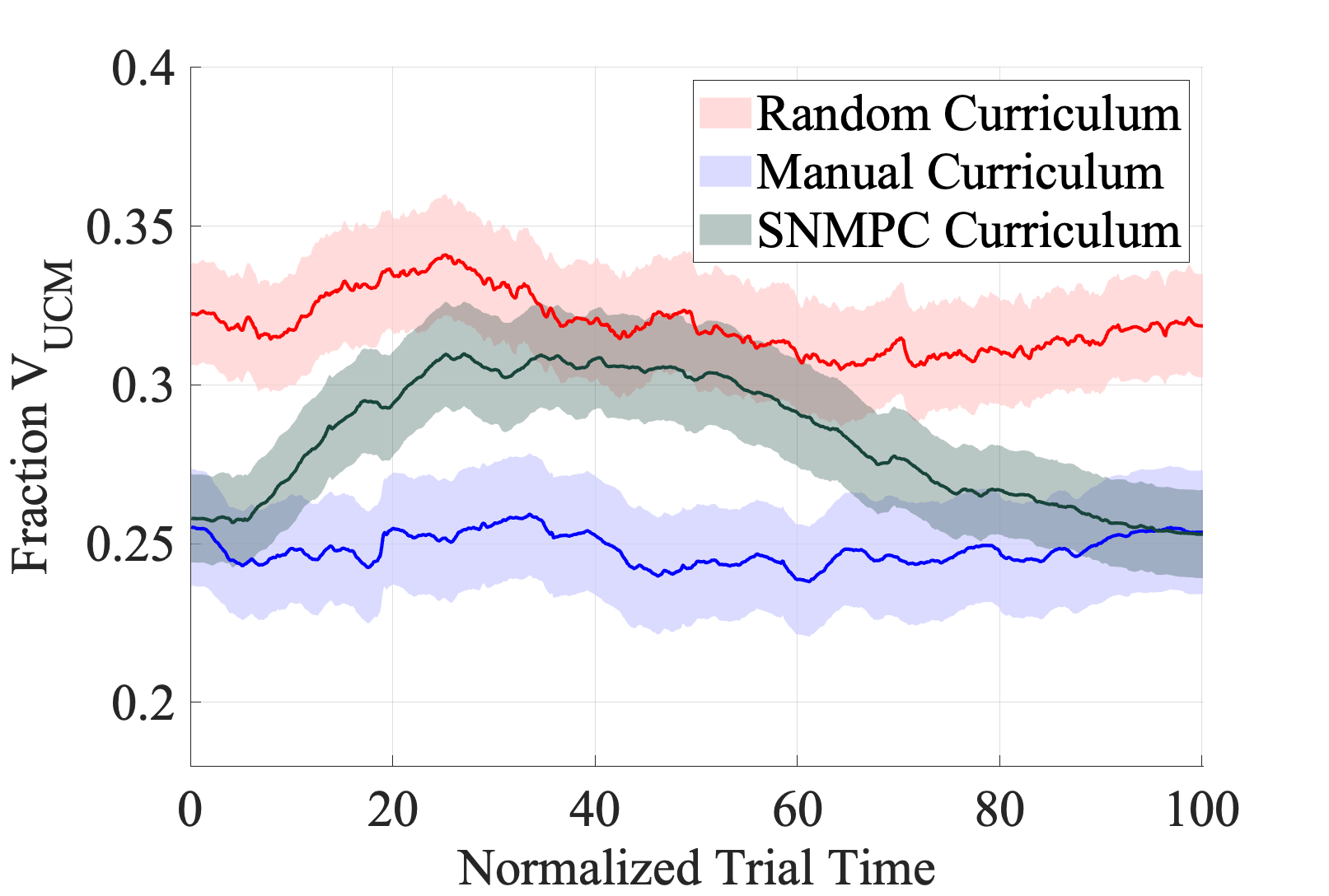}
        \label{fig:ucmp1}
    \end{subfigure}\hspace{-1.2em}%
    ~
    \centering
    \begin{subfigure}{0.34\linewidth}
	    \centering
        \caption{}
        \includegraphics[width=1\linewidth, height=1.1\linewidth, keepaspectratio]{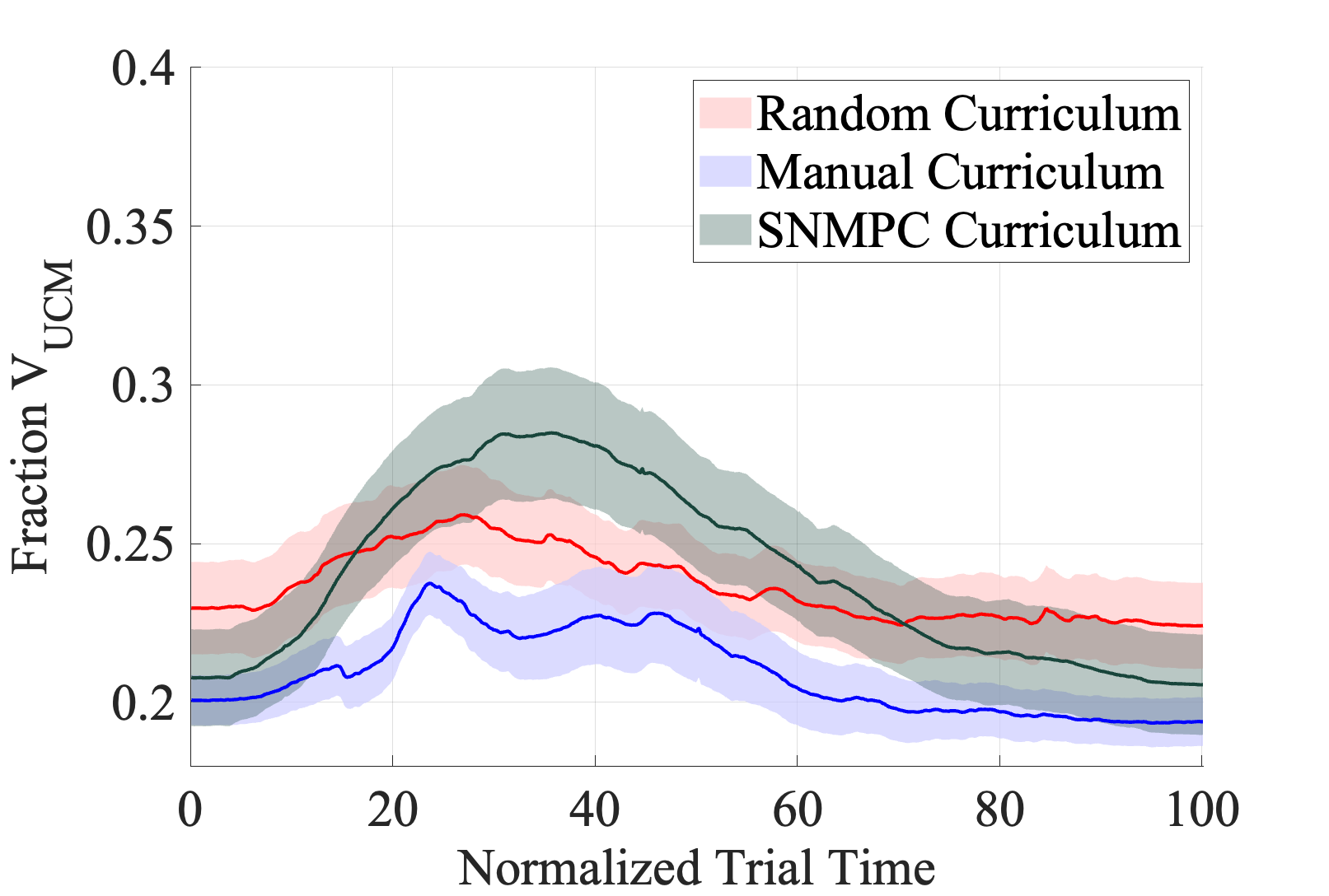}
        \label{fig:ucmp2}
    \end{subfigure}\hspace{-1.2em}%
    ~
    \centering
    \begin{subfigure}{0.34\linewidth}
	    \centering
        \caption{}
        \includegraphics[width=1\linewidth, height=1.1\linewidth, keepaspectratio]{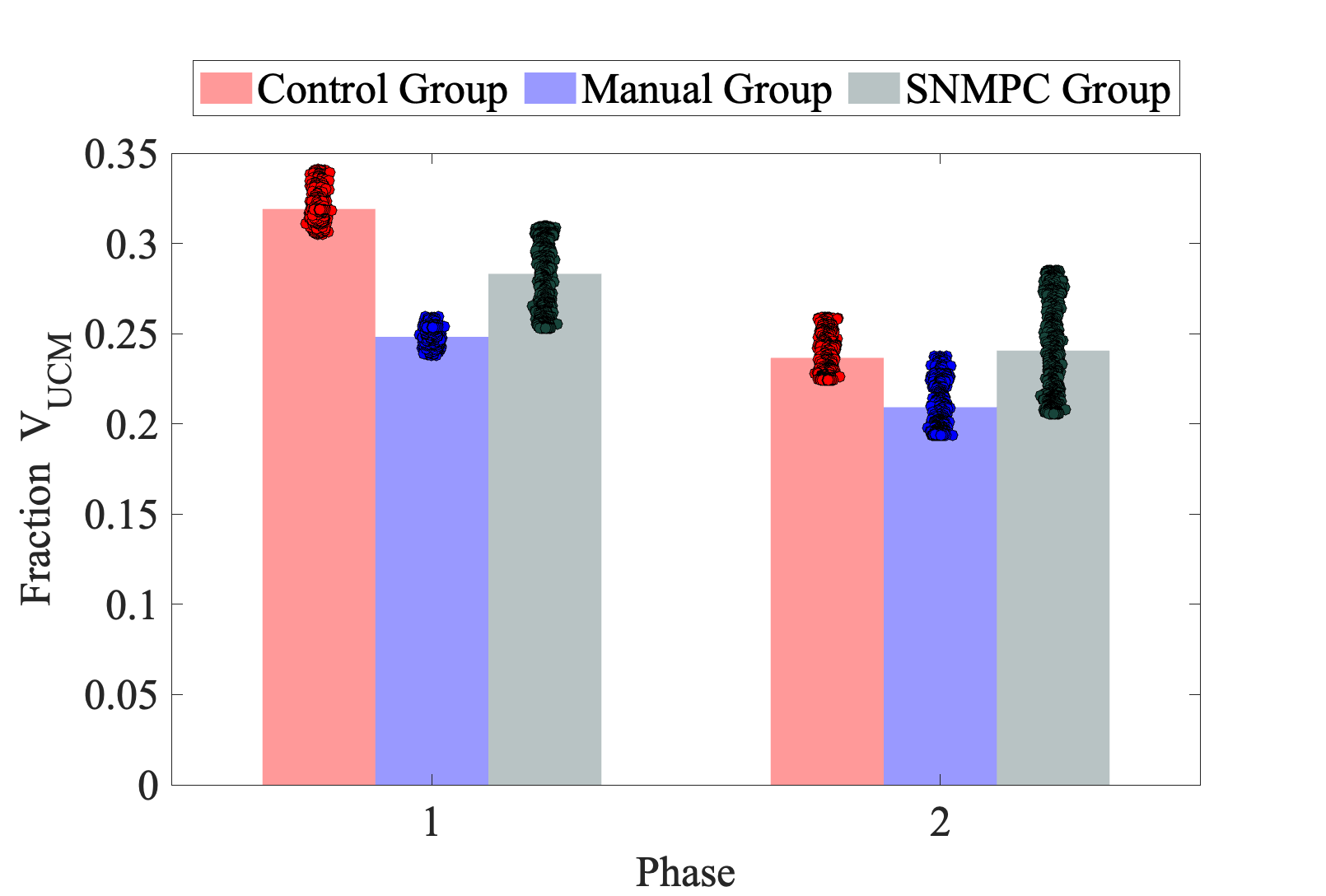}
        \label{fig:ucm-p1p2}
    \end{subfigure}\hspace{-1.2em}%
    \vspace{-0.15in}
    \caption{\textbf{Uncontrolled Manifold Analysis:} The fraction of trial-to-trial variability in the hand finger-joints of the participants projected onto the UCM of the task for (a) phase $1$, (b) phase $2$ as a function of the normalized trial time, and (c) the total fraction of $\subscr{V}{UCM}$ across the trials for the two phases under the three sequencing strategies shows that the variability decreases for all groups, an evidence of task learning. Additionally, the SNMPC group allocates a larger total fraction of their variability to the UCM than the control and manual groups in phase $2$, implying a smaller fraction of variability in task space, indicating better skill learning. Furthermore, $\subscr{V}{UCM}$ exhibits a bell-shaped curve in phase $2$, with the SNMPC group exhibiting the highest peak, corroborating expedited skill acquisition. Phase $1$: training blocks $1$-$3$. Phase $2$: training blocks $4$-$6$.}
    \label{fig:ucm}
    \vspace{-0.15in}
\end{figure*}
\subsubsection{Task Performance}
Smoothness of movement-based metrics has been shown to effectively capture improvements in motor learning during rehabilitation~\cite{lu2021evaluating, sungeelee2024interactive}. Thus, Straightness of Trajectory \texttt{SoT} (defined in Section \ref{sec:manualTS}) was employed as the performance metric to examine the task performance of the three groups on our novel motor task; a lower \texttt{SoT} value is desirable.
\begin{table}[b!]
    \vspace{-0.2in}
    \centering
    \caption{\textbf{Two-tailed Linear Mixed Model contrasts for group differences.} The p-values for the hypothesis that the groups have different \texttt{SoT} values across trials show that the SNMPC group (SG) achieves lower \texttt{SoT} values compared to control (CG) and manual (MG) groups, implying better performance throughout the gameplay. The SNMPC group also requires fewer trials than the control group to achieve low values of \texttt{SoT} thresholds, indicating expedited performance convergence.
    $(\cdot)^*$ are the p-values for reverse comparison.}
    \begin{tabular}{||c|c||c|c|c||}
        \hline
        \multicolumn{2}{||c||}{\multirow{2}{*}{}}&
        \multicolumn{3}{c||}{\textbf{Two-tailed p-value for}} \\
        \cline{3-5}
        \multicolumn{2}{||c||}{}& \textbf{SG $<$ CG} & \textbf{SG $<$ MG} & \textbf{MG $<$ CG} \\
        \hline \hline
        \multicolumn{2}{||c||}{\textbf{SoT Values}} & \textbf{$<$0.0001} & \textbf{$<$0.0001} & \textbf{0.0001$^*$} \\
        \hline \hline
        \multirow{4}{*}{\shortstack{\textbf{SoT} \\ \textbf{Thresholds}}}  & 0.24 & \textbf{0.0216} & 0.1150 & 0.4657\\
                                        \cline{2-5}
                                        & 0.29 & 0.3302 & 0.3025 & 0.9541$^*$\\
                                        \cline{2-5}
                                        & 0.34 & 0.2169 & 0.7033 & 0.3924\\
                                             \cline{2-5}
                                        & 0.39 & 0.2576 & 0.5184 & 0.6260\\
        \hline
    \end{tabular}
    \label{tab:SOT_pVal}
    \vspace{-0.25in}
\end{table}
Fig. \ref{fig:SOT_curves} shows the evolution of \texttt{SoT} for control, manual, and SNMPC groups across the game trials, and Fig. \ref{fig:SOT_threshold} shows the distribution of the number of trials required for different target pairs in the three groups to converge to the corresponding \texttt{SoT} values.
The idea behind looking at the mean number of trials (across participants and target pairs) required to achieve specific \texttt{SoT} thresholds is that a curriculum design dictates the target pair selections, prescribing easier targets less frequently than challenging ones to accelerate performance/skill improvement. 
Further, an adaptive and personalized curriculum design could help achieve even faster \texttt{SoT} convergence by carefully selecting the target pairs conducive to learning relevant finger motions.
Table \ref{tab:SOT_pVal} shows the p-values comparing the performance of different groups, where the first row reports results for the whole \texttt{SoT} curves in Fig. \ref{fig:SOT_curves}, and the next four rows contain results for \texttt{SoT} thresholds in Fig. \ref{fig:SOT_threshold}.

The results show a decrease in the \texttt{SoT} values across trials, and the mean number of trials to achieve a desired \texttt{SoT} also decreases with the increasing \texttt{SoT} threshold. The p-values in Table \ref{tab:SOT_pVal} show that the participants in the SNMPC group have statistically significantly lower \texttt{SoT} values compared to other groups across all the game trials. Additionally, the SNMPC group requires statistically significantly fewer trials than the control group on average ($\sim79$ versus $\sim50$ trials) to achieve an \texttt{SoT} threshold of $0.24$. A possible explanation for the decrease in mean differences between the groups as \texttt{SoT} thresholds increase could be that SNMPC and manual group participants train on harder targets initially.
\begin{figure*}[t!]
    \centering
    \begin{subfigure}{0.55\linewidth}
	    \centering
        \caption{}
        \includegraphics[width=1\linewidth, keepaspectratio]{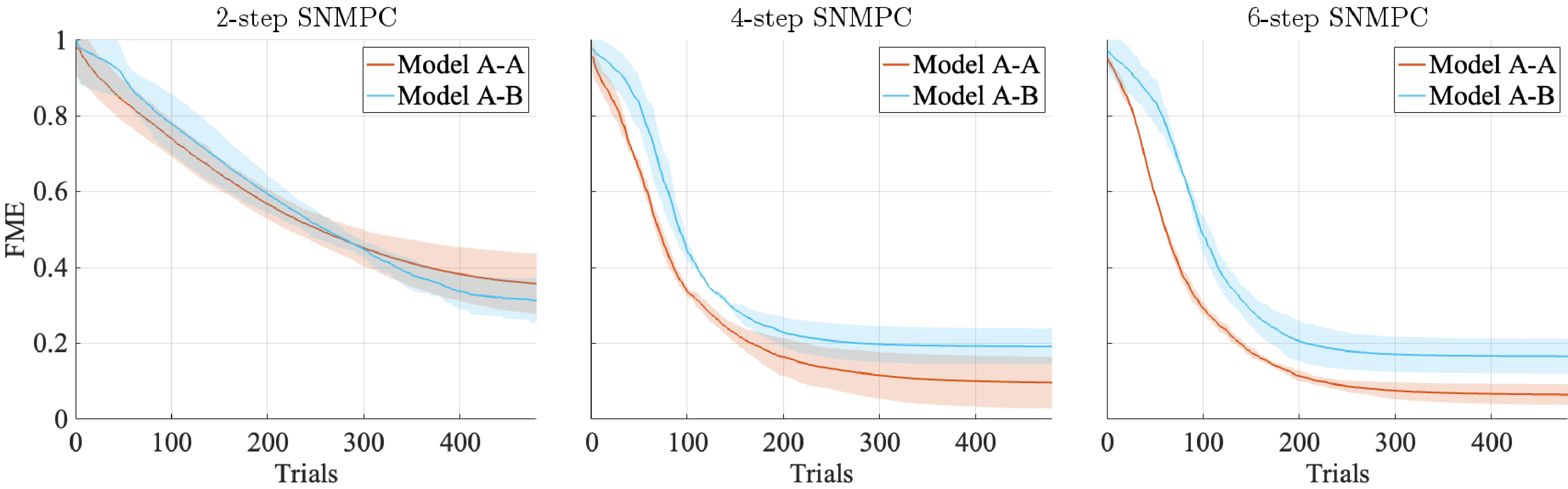}
        \label{fig:SNMPC_robustness}
    \end{subfigure}
    ~
    \begin{subfigure}{0.215\linewidth}
	    \centering
        \caption{}
        \includegraphics[width=1\linewidth, keepaspectratio]{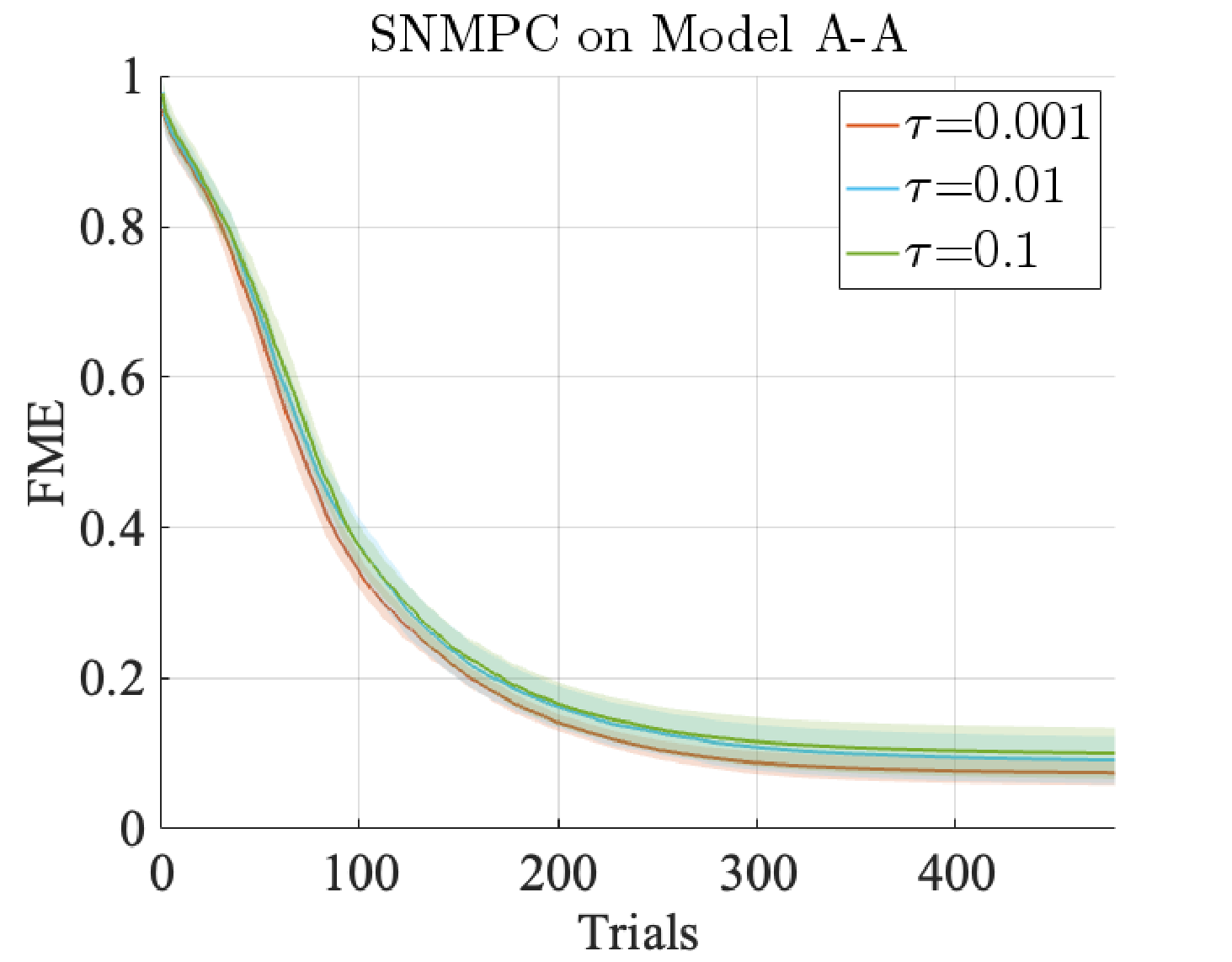}
        \label{fig:tau_effect}
    \end{subfigure} \hspace{-1.7em}%
    ~
    \begin{subfigure}{0.21\linewidth}
	    \centering
        \caption{}
        \includegraphics[width=1\linewidth, keepaspectratio]{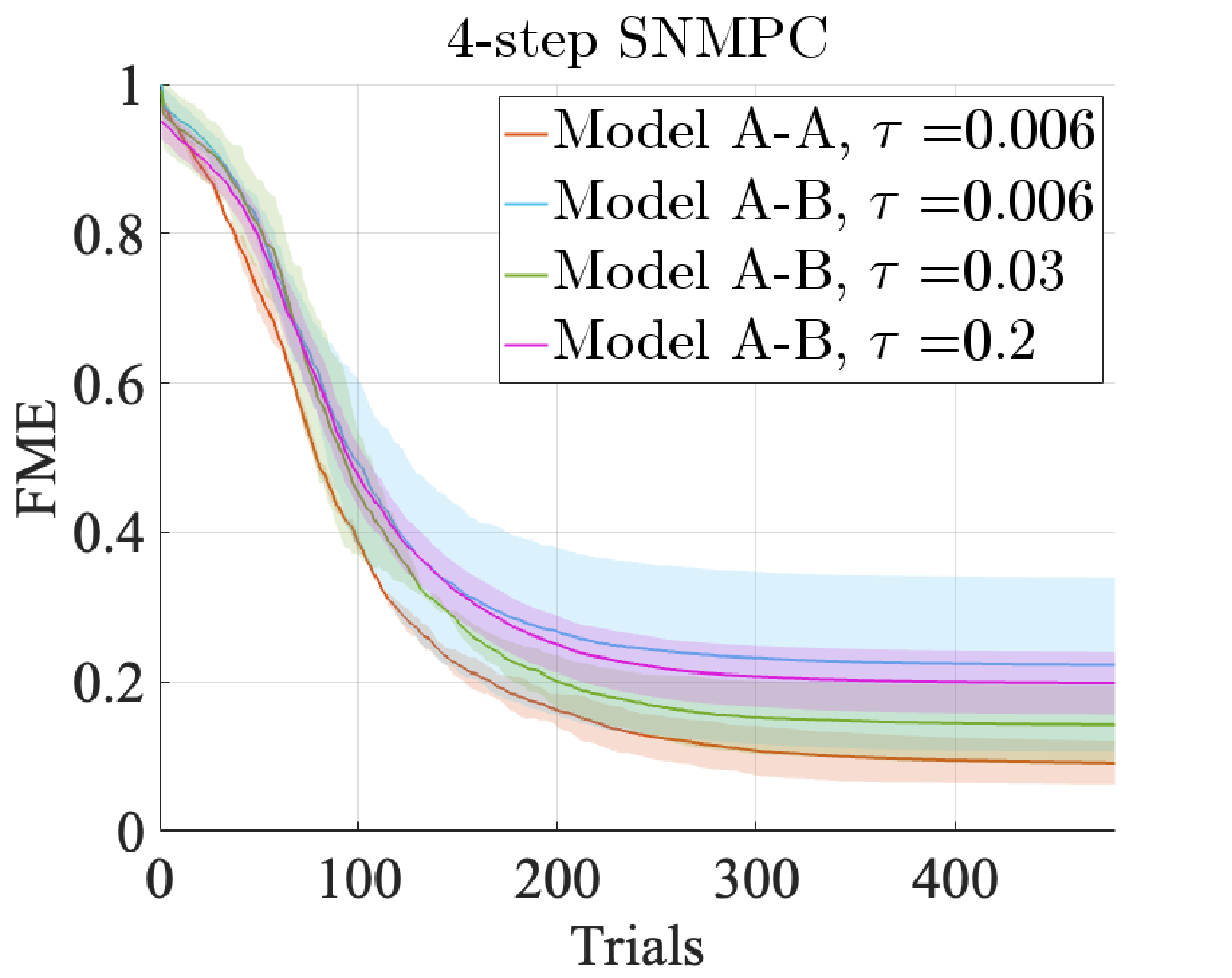}
        \label{fig:tau_robustness}
    \end{subfigure}
    \vspace{-0.15in}
    \caption{\textbf{Robustness of SNMPC Curriculum to Model Mismatches:} \texttt{FME} curves for fitted participant model A playing the game on SNMPC curriculum, with SNMPC using model A (A-A) or using model B (A-B), show that learning performance degrades (a) for model A-B with increasing lookahead horizon ($2,4,6$), indicating the importance of the correct model in SNMPC. (b) In the case of no model mismatch (Model A-A), increasing $\tau$ deteriorates learning. (c) However, in the presence of model mismatch, $\tau$ helps improve learning, with the lowest learning error achieved for $\tau = 0.03$, signifying the importance of $\tau$ against inevitable model mismatches. $\tau$ modulates the level of controlled stochasticity in policy generation~\eqref{eq:policy_prob}.}
    \label{fig:robustness}
    \vspace{-0.25in}
\end{figure*}
\subsubsection{Uncontrolled Manifold Analysis}
We now investigate the underlying structure of motor coordination for the three experiment groups. To understand how the participants manage the high motor redundancy in our motor task, we leverage the uncontrolled manifold (UCM) analysis \cite{scholz1999uncontrolled, scholz2000identifying}.
We consider finger joint angles as elemental variables, screen cursor positions as performance variables, thus giving matrix $C$ as the Jacobian relating the two variables. $\subscr{V}{UCM}$ represents variability in joint angles lying in the uncontrolled manifold (the null space); $\subscr{V}{ORT}$ denotes the variability in the orthogonal subspace (the task space).

Results from the total fraction of $\subscr{V}{UCM}$ throughout the normalized trials for each group across the two phases in Fig. \ref{fig:ucm-p1p2} show a statistically significant $(F_{(1,33)} = 316.76, p<0.0001)$ decrease in variability across the phases for all groups. The control group has the highest fraction of $\subscr{V}{UCM}$ among all three groups initially. However, as the task and motor learning progress, the SNMPC group exhibits the highest fraction of its total variability within the UCM in phase $2$, surpassing both control and manual groups, as shown in Fig. \ref{fig:ucm-p1p2}. $\subscr{V}{UCM}$ exhibits a unimodal, bell-shaped profile across the normalized trial time, particularly in phase $2$ for all groups, with SNMPC group exhibiting the highest unimodal peak.

\section{Discussion} \label{sec:discussion}

This study demonstrates that an automated, model-informed adaptive curriculum significantly accelerates skill acquisition in novel, high-dimensional motor tasks. While structured heuristics outperform random practice, personalizing the curriculum to the learner's latent skill state yields the most substantial gains in performance and learning efficiency. Ultimately, this underscores the limitations of purely performance-based metrics for complex, high-DoF tasks and highlights the value of model-informed optimization. 

\subsection{Structured Practice Accelerates Learning, but Heuristics Have Limits}
Consistent with prior work in learner-adaptive training \cite{choi2008performance, simon2008win}, our results confirm that structured practice is more effective than random task sequencing. Participants in both the manual heuristic and SNMPC groups required significantly fewer trials to improve their \texttt{SoT} and showed faster convergence of learning error compared to the control group. This establishes a clear benefit for moving beyond random practice in motor learning paradigms. 
However, the efficacy of a curriculum hinges on the metrics used to guide it.
While many studies have shown the benefits of curriculum-based practice for low-DoF tasks where performance metrics are a reliable indication of skill~\cite{wadden2018individualized}, in high-dimensional tasks with significant motor redundancy, the link between observable performance and true skill acquisition may be weak or even misleading \cite{guadagnoli2004challenge}. This challenge complicates the design of effective heuristic-based curricula, which rely on an expert's ability to select appropriate metrics, a non-trivial task for complex motor systems. This is reflected in our UCM analysis, where the manual group participants have consistently higher fraction of trial-to-trial variability in the task space $\subscr{V}{ORT}$ ($=1-\subscr{V}{UCM}$). This is indicative of inferior skill learning~\cite{latash2007toward} and task performance, as fraction $\subscr{V}{ORT}$ is correlated with the participants' ability to identify and utilize the task space to achieve desired objectives. These findings are complemented by our \texttt{SoT} results.
\subsection{Model-Based Adaptation is Key for High-Dimensional Skill Acquisition}
To overcome the limitations of heuristics, our work shows the clear superiority of a curriculum that adapts to the real-time estimate and evolution of the learner's skill state and its impact on the performance metrics. The SNMPC algorithm, by integrating a computational model of motor learning, consistently outperformed the performance heuristics-based curriculum in both task performance and learning speed.
The advantage of this model-based approach is its ability to look beyond observed performance and leverage its dynamics to estimate and optimize over the latent skill state and performance evolution rather than over past observable outcomes.
Consistent with this claim, UCM analysis indicates how SNMPC group participants can learn and utilize the forward mapping to achieve the task goals faster than other groups, with the highest trial-to-trial variability in the UCM in the later phase of the task. We also observe a unimodal, bell-shaped $\subscr{V}{UCM}$ profile, indicating feedback-based movement corrections at the start and end of the trials, and feedforward motions based on participants' estimate of the forward mapping matrix during the trial~\cite{goodman2006feed}. Additionally, the peak is highest for SNMPC group participants on average, supporting the expedited skill learning results for the SNMPC-based curriculum from previous analyses. 
It is noteworthy to mention that the control group participants start with a much higher fraction of their total variability in the UCM at normalized trial time $0$ in phase $1$ compared to the other two groups. This could be a result of manual and SNMPC groups training on harder targets initially.

\subsection{Practical Implementation and Robustness of SNMPC}   \label{sec:discussion:MPC_robustness}
In contrast to~\cite{kamboj2025expediting}, we successfully deploy the SNMPC using a population-average HML model combined with personalized skill estimates. This eliminates the overhead of a separate model-fitting session and prevents early-stage skill carry-over. Our framework proves effective even without a fully personalized model, illustrating a practical trade-off between model fidelity and experimental complexity.

The feedback structure of the receding-horizon controller provides a degree of robustness to potential model mismatches. The stochasticity parameter $\tau$ is also critical in managing this trade-off. Our simulation results (Fig. \ref{fig:SNMPC_robustness}) confirm that model mismatch degrades skill learning, exacerbating as the prediction horizon lengthens, since the SNMPC relies more on the incorrect model. While a lower $\tau$ leads to a more optimal policy and faster learning with an accurate model (Fig. \ref{fig:tau_effect}), this reliance on the model becomes a liability in the presence of model uncertainties. As hypothesized, introducing controlled stochasticity with a moderately high $\tau$ improves learning performance when the model is not perfectly aligned with the participant (Fig. \ref{fig:tau_robustness}), offsetting the algorithm's over-reliance on the inaccurate model. However, setting $\tau$ too high is also detrimental, as the policy deviates too far from optimal. Thus, $\tau$ is a key parameter for tuning the balance between optimality and robustness to the inevitable model mismatches when using a generalized model. Future work could examine its impact on learning outcomes in an experimental setting.

\subsection{Future Directions}
The development of automated curriculum design has significant implications for high-dimensional motor learning and rehabilitation. For example, combining our SNMPC framework with assist-as-needed controllers \cite{castiblanco2021assist, agarwal2017subject} merges adaptive task-sequencing with physical assistance, creating optimized training regimens with minimal expert supervision. We view this framework as part of an ongoing evolution in personalized human-machine interfaces. While prior systems focused on best-parameter identification~\cite{patton2004robot} or used Bayesian human-in-the-loop optimization to maximize information at minimal cost~\cite{zhang2017human}, our approach advances the field by explicitly controlling task difficulty, taking into account the user's skill level via a population-average model. To eventually achieve a fully closed-loop learning design, it is critical to move beyond fixed learner models~\cite{broad2018learning, slade2024human}. A powerful next step is co-adapting the HML model in real-time as the participant trains. Continuous parameter estimation could progressively personalize the curriculum to individual learning dynamics and strategy shifts. This co-evolution of curriculum and learner model promises even faster and more robust skill acquisition.
Additionally, this framework can be extended to broader learning paradigms using other high-DoF motor systems.

\section{Methods} \label{sec:materialandmethods}
\subsection{Experimental Details}
To study the effect of proposed scheduling algorithms, participants were divided into three groups playing the target capture game with random scheduling (control group), manual scheduling (manual group), and SNMPC-based scheduling (SNMPC group). 
A total of $36$ healthy participants (age $19.19\pm 1.28$ years) were recruited for this study, with $12$ in each experiment group, and informed consent was obtained from each participant\footnote{The human behavioral experiments were approved under Michigan State University Institutional Review Board Study ID LEGACY14-431M.}.
We ran experiments to assess the efficacy of the proposed SNMPC-based target sequencing algorithm in expediting motor learning in healthy human participants compared to the other two curriculum design approaches.
All $12$ control group participants played the game first, and their gameplay data were used to fit the HML model to each participant. The $12$ HML model parameter sets were averaged to obtain a \emph{mean} HML model, which was used in skill state estimation and the SNMPC algorithm for the SNMPC group. This is in contrast to our previous work \cite{kamboj2025expediting}, which required an additional prior session for each participant to fit individualized models for state estimation and curriculum design. Interested readers can refer to Appendix for more details on the HML model, the parameter fitting procedure, and the fits.

\subsection{Statistical Analysis}
Statistical inference for task performance is performed using a Linear Mixed Model (LMM) with repeated measures of performance metrics for each participant, accounting for across and within-subject correlations. The LMM parameters are estimated using Restricted Maximum Likelihood~\cite{corbeil1976restricted}. For comparing group task performances based on \texttt{SoT}, we perform two analyses. First, we look at the \texttt{SoT} values for participants in the three groups across the game trials. Second, we compute how many game trials each participant group takes to converge to a particular (\texttt{SoT}) threshold under the three curricula across different target pairs. \textit{Post hoc} two-tailed contrasts tested the hypotheses that the \texttt{SoT} values and mean number of trials to achieve the thresholds are different for the SNMPC group compared to the control and manual groups, employing a Bonferroni correction for the two comparisons~\cite{wright1992adjusted}.

\subsection{Uncontrolled Manifold Analysis}
The UCM framework partitions movement variability into two functionally distinct components: $\subscr{V}{UCM}$, desirable finger-joint variability within the task's null space that does not affect cursor position, and $\subscr{V}{ORT}$, undesirable variability orthogonal to the UCM that alters performance variables. For trial-to-trial analysis, trials were grouped by target pair and time-normalized ($0.1\%$ increments). Deviations from the mean joint-angle trajectory, representing the stabilized desired reference, were projected onto the null space and its orthogonal complement via the mapping matrix $C$. These projections were squared and normalized by their respective DoFs and trial counts. To ensure sufficient sample sizes, data were analyzed across Phase $1$ (training blocks $1$–$3$) and Phase $2$ (blocks $4$–$6$). We report the fraction of $\subscr{V}{UCM}$ to total variance (where $\subscr{V}{ORT}=1-\subscr{V}{UCM}$) in Fig.~\ref{fig:ucm}. Participant-level fractions were averaged across target pairs to compute group means and $95\%$ confidence intervals (Fig. \ref{fig:ucmp1}, \ref{fig:ucmp2}), which were evaluated using a $2\times 3$ (Phase × Group) mixed-model ANOVA.

\bibliographystyle{IEEEtran}
\bibliography{mybib, motor_learning, masterbib}

\appendix
\section*{Appendix}
\section{Trial-by-trial HML Model}
The HML model proposed in our previous work \cite{kamboj2024human} is summarized below
\begin{subequations}    \label{appendix:HML_model}
    \begin{align}
        \dot{\bs x} =&\ C \bs u  \label{HML_model:x} \\
        \dot{\delta \bs q} =& -a \delta \bs q + {\bs u} + \bs \xi_q \label{appendix:HML_model:deltaq}\\
        \dot{\hat{W}} =& -\gamma \tilde{W} \Phi \delta \bs q \delta \bs q\tran \Phi\tran \label{appendix:HML_model:W_hat} \\
        \dot{ {\bs u}} =& - \eta \left( \left(\Phi\tran \hat{W}\tran \hat{W} \Phi + \mu I \right){{\bs u}} - k_P \Phi\tran \hat{W}\tran (\supscr{\bs x}{des} - \bs x) \right) + \bs \xi_u \label{appendix:HML_model:u} \\
        \dot{\bs q} =&\ \bs u.
    \end{align}
\end{subequations}
We denote $\hat{C}(t)$ as the participant's implicit estimate of the forward mapping matrix $C$ at time $t$. Using the synergies extracted from human participant data, $\Phi$, we factor $\hat{C} = \hat{W}\Phi$, where $\hat{W}$ is called the parameter estimate matrix, which captures the estimated weights that the human participant assigns to each synergy.
$\tilde{W} = \hat{W} - W$ is thus the parameter estimation error. $\delta \bs q$ is the filtered increment in joint angles governed by the dynamics in \eqref{appendix:HML_model:deltaq}, $\bs \xi_q$ is the perceptual noise that captures the inaccuracies in human motion perception, $\supscr{\bs x}{des}$ is the desired (target point) cursor position, and $\bs \xi_u$ is the exploratory noise that captures the exploration by humans in the joint velocity space, and together they give us the system/process noise $\bs \omega(t)$.
The following are the HML model parameters: $\gamma$ is the forward learning rate, $\eta$ is the inverse learning rate, $\mu$ is the optimality parameter, $k_P$ is the control intensity, $\sigma_u$ is the exploratory noise intensity, $\sigma_q$ is the perceptual noise intensity, and $a$ is the perceptual recency parameter.

The curriculum design problem for our target capture game is posed as the selection of a sequence of target points for each trial. Since the target point is fixed for each trial, we rewrite our system dynamics as \emph{trial-by-trial dynamics} to aid in curriculum algorithm design. Although HML model equations describe the intra-trial time evolution of system states, we are only concerned with the states at the end of a trial. Thus, consider the stochastic discrete nonlinear system
\begin{align}
    {\bs \zeta}_{j+1} &= \Psi(\bs \zeta_j, \bs \upsilon_{j+1}, \Omega_j), \\
    \bs y_j &= \Theta(\bs \zeta_j, \rm N_j),
\end{align}
that captures the evolution of the states of human motor learning from the end of trial $j$ to the end of trial $j+1$ of the target capture gameplay.
Here, $j \in \naturals$, $\bs \zeta_j = \bs \zeta(T_j)$ is the system state at the end of trial $j$ ending at time $T_j$, $\bs \upsilon_{j+1} \in \mb U \subset \real^{n}$ is the target point for trial $j+1$, $\mb U$ is a non-empty, finite and countable set of admissible target points, $\bs y_j = \bs y(T_j)$ is the system output, and $\Psi$ and $\Theta$ are nonlinear functions describing the trial-to-trial system dynamics and outputs, respectively. For a trial $j$ starting at $t_j^s$ and ending at $t_j^e$, $\Omega_j$ is a function of the system noise $\bs \omega(t), t\in  [t_j^s, t_j^e]$, and $\bs \rm N_j$ is a function of $\bs \nu_j(t), t\in [t_j^s, t_j^e]$, the measurement noise for trial $j$. The system noise $\bs \omega$ and the measurement noise $\bs \nu$ are modeled as white noise processes.
\section{Fitting HML Model to Human Participant Data}
\rev{To obtain the HML model parameters, we use the human participant experiment data from the control group (training on random target sequences), since a broad excitation of the system generally helps obtain better model fits. However, this is not required since the human motor system is inherently noisy; data from any other group can be used to obtain a population-averaged model of human motor learning.}
The task performance is quantified by three metrics - reaching error (\texttt{RE}), straightness of the trajectory (\texttt{SoT}), and the trajectory error (\texttt{TE}), to ascertain the performance of the HML model while fitting the data. Reaching error (\texttt{RE}) in each trial in the human participant data is calculated as the Euclidean distance of the cursor from the target point, either $2$s after the movement starts or at the end of the movement, whichever is earlier.
The straightness of the trajectory (\texttt{SoT}) is defined as an aspect ratio of the maximum perpendicular distance of the trajectory from the straight line joining the start and end points, to the straight line distance between the start and the end points. The trajectory error (\texttt{TE}) is simply the norm difference between the cursor trajectories from the data and the cursor trajectories from the HML model.
Owing to the stochastic non-linearity of the proposed HML model, we use the multi-objective optimization genetic algorithm NSGA-II~\cite{deb2002fast} to find the optimal parameters over the parameter space using $\subscr{f}{\texttt{RE}}=\norm{\subscr{\texttt{RE}}{model} - \subscr{ \texttt{RE}}{data} }_2$, $\subscr{f}{\texttt{SoT}}=\norm{\subscr{\texttt{SoT}}{model} - \subscr{\texttt{SoT}}{data}}_2$, and $\subscr{f}{\texttt{TE}}=\norm{\texttt{TE}}_2$ as the three objectives.
The subscripts denote whether the metric is formed using experimental data or the HML model. The duration of each trial of the HML model was consistent with the human experiment data, providing a fair basis for objective function calculation. NSGA-II was run in two phases: exploration and exploitation. First, the exploration phase is run for $250$ generations using the simulated binary crossover (SBX) operator with probability $0.9$ and eta (degree of similarity) value of $15$, and polynomial mutation (PM) operator with probability $0.17$ and eta value of $10$, over a population size of $256$. This is followed by the exploitation phase, which starts with a population comprised of the non-dominated front members from the exploration phase population using their non-dominated ranking and crowding distance. This phase is run for another $250$ generations using SBX with probability $0.9$ and eta $20$, and PM with probability $0.1$ and eta $15$, over a population size of $128$. Additionally, after every $5$ generations of the exploitation phase, we update the objective values of the non-dominated front members by resampling from the HML model $3$ times to account for stochasticity in the proposed HML model. For every generation of each phase, the number of offsprings are selected to be the same as the population size.
Out of the $\min\{\subscr{f}{\texttt{RE}}\}$ fits selected from $10$ runs of NSGA-II, the parameter fits with minimum $\subscr{f}{\texttt{RE}}$ and $\subscr{f}{\texttt{SoT}} < \textup{avg}\{\subscr{f}{\texttt{SoT}}\}$ are chosen for a particular subject.
Perceptual recency parameter $a$ was heuristically chosen to be a large value $\sim O(10)$, and the target size $\rho_x$ value was chosen the same as in the experiments. The fitted model parameter values used in simulation studies are
\begin{align}
    [\gamma, &\eta, \mu, k_P, \sigma_u, \sigma_q, a] = [0.0001, 0.6348, 5.7399, 11.3432, 0.8358, 0.0010]. \notag
\end{align}
\section{Gameplay Trajectories}
\begin{figure*}[h!]
    \centering
    \begin{subfigure}{0.3\linewidth}
	    \centering
        \caption{}
        \includegraphics[width=1\linewidth, height=1\linewidth, keepaspectratio]{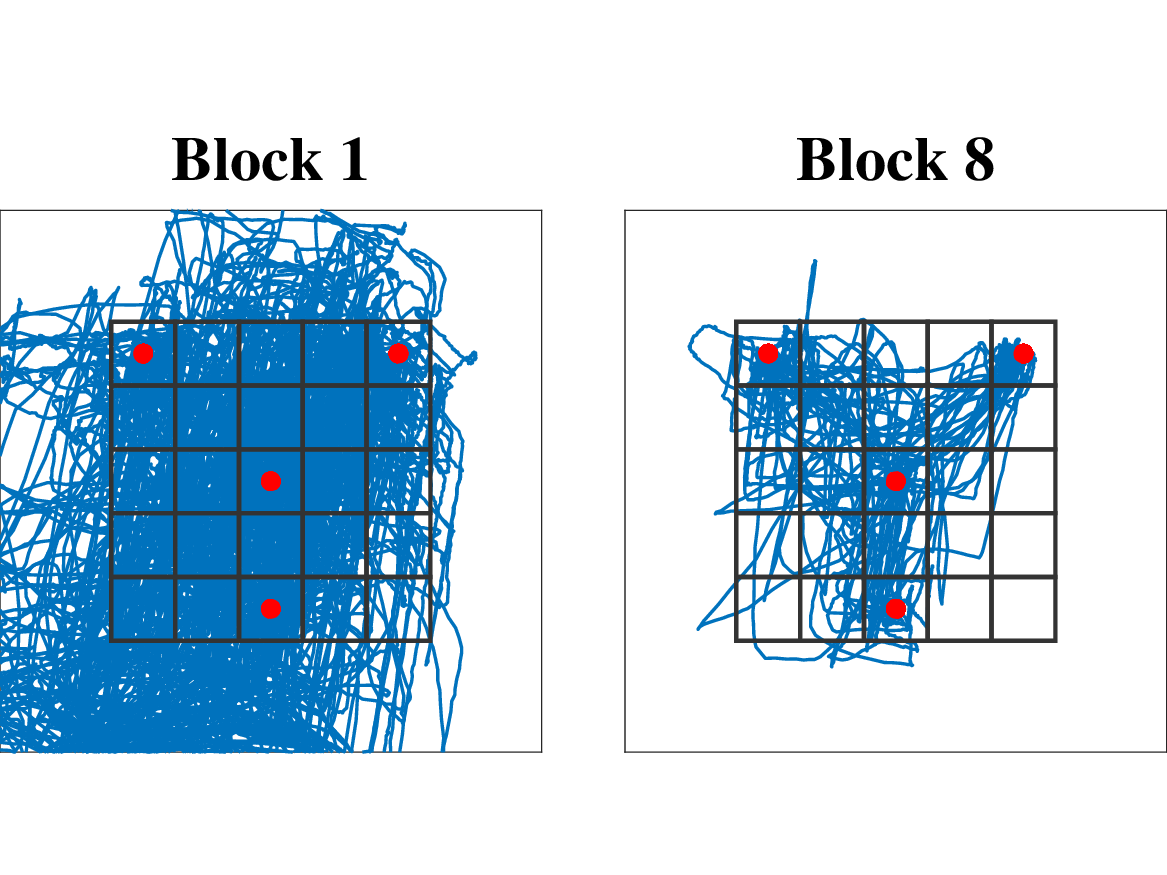}
    \end{subfigure}
    ~
    \centering
    \begin{subfigure}{0.3\linewidth}
	    \centering
        \caption{}
        \includegraphics[width=1\linewidth, height=1\linewidth, keepaspectratio]{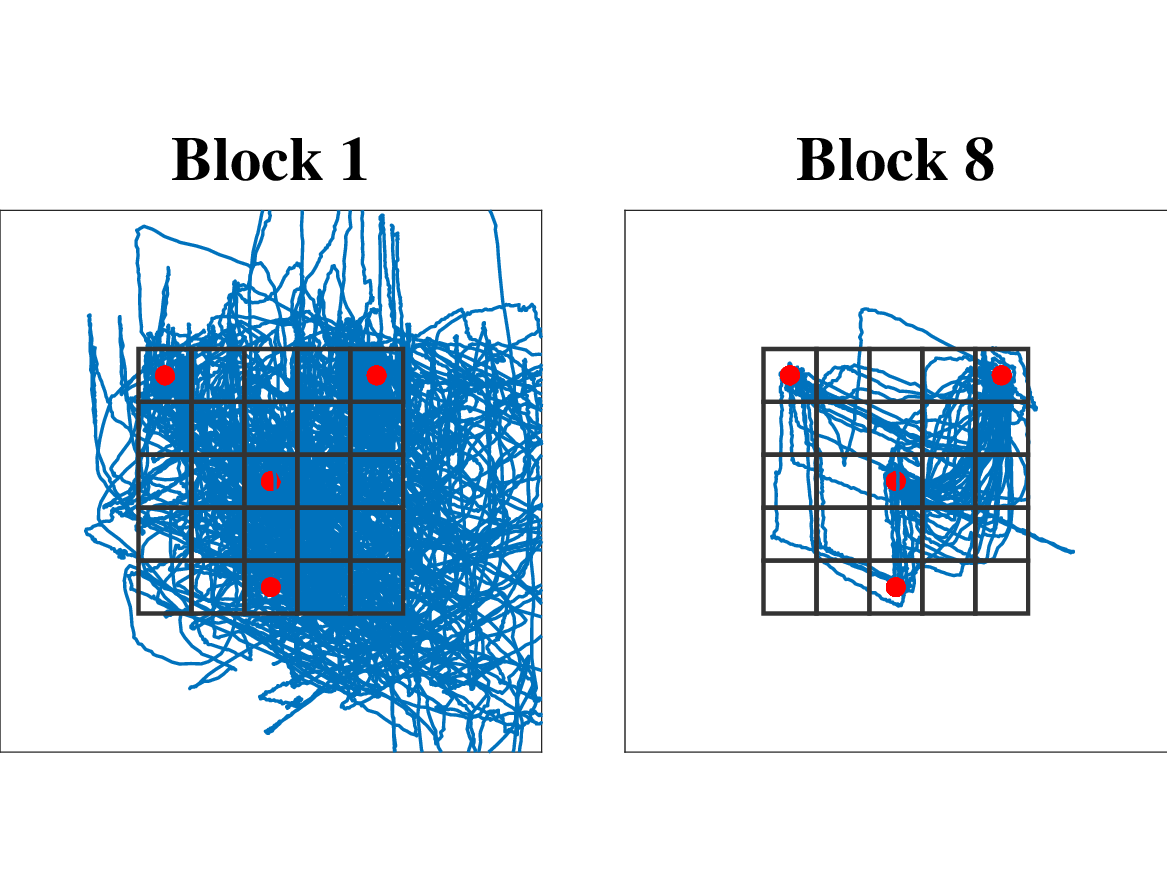}
    \end{subfigure}
    ~
    \centering
    \begin{subfigure}{0.3\linewidth}
	    \centering
        \caption{}
        \includegraphics[width=1\linewidth, height=1\linewidth, keepaspectratio]{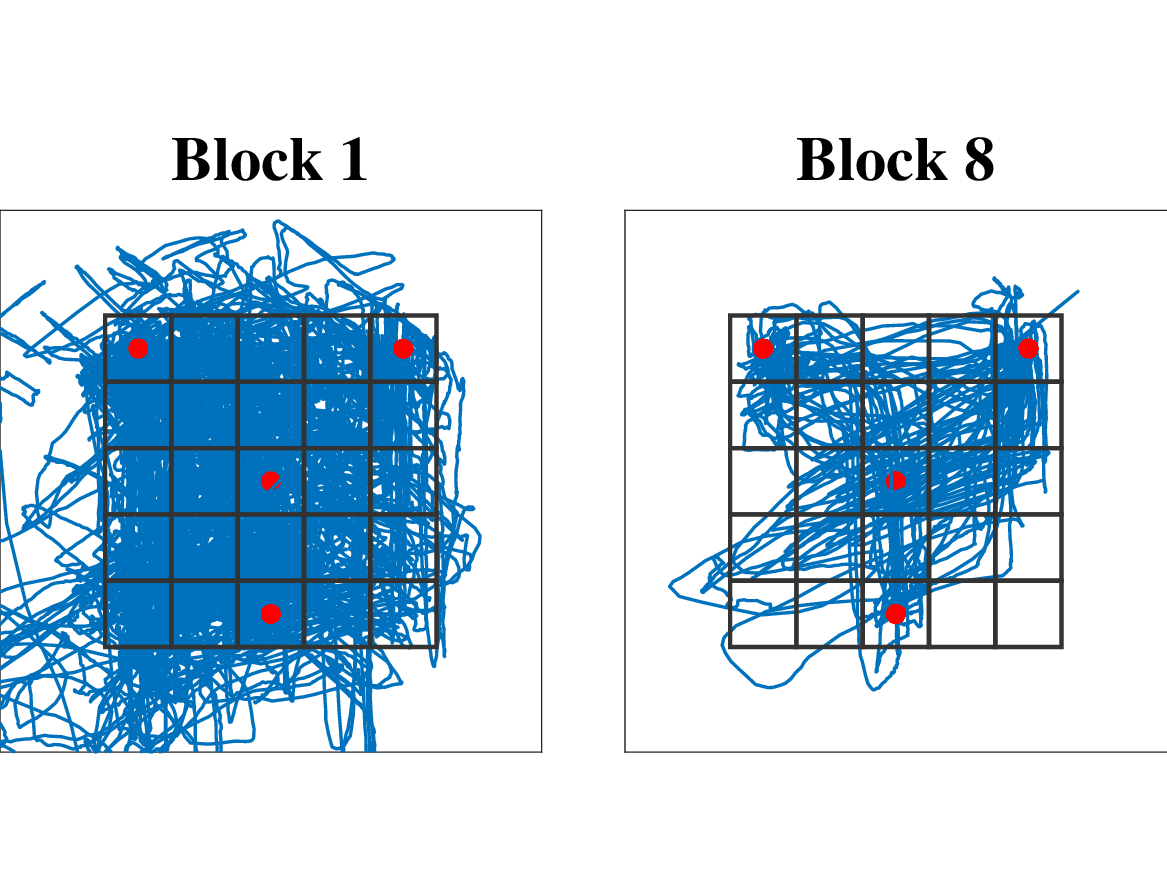}
    \end{subfigure}
    \\
    \begin{subfigure}{0.3\linewidth}
	    \centering
        \caption{}
        \includegraphics[width=1\linewidth, height=1\linewidth, keepaspectratio]{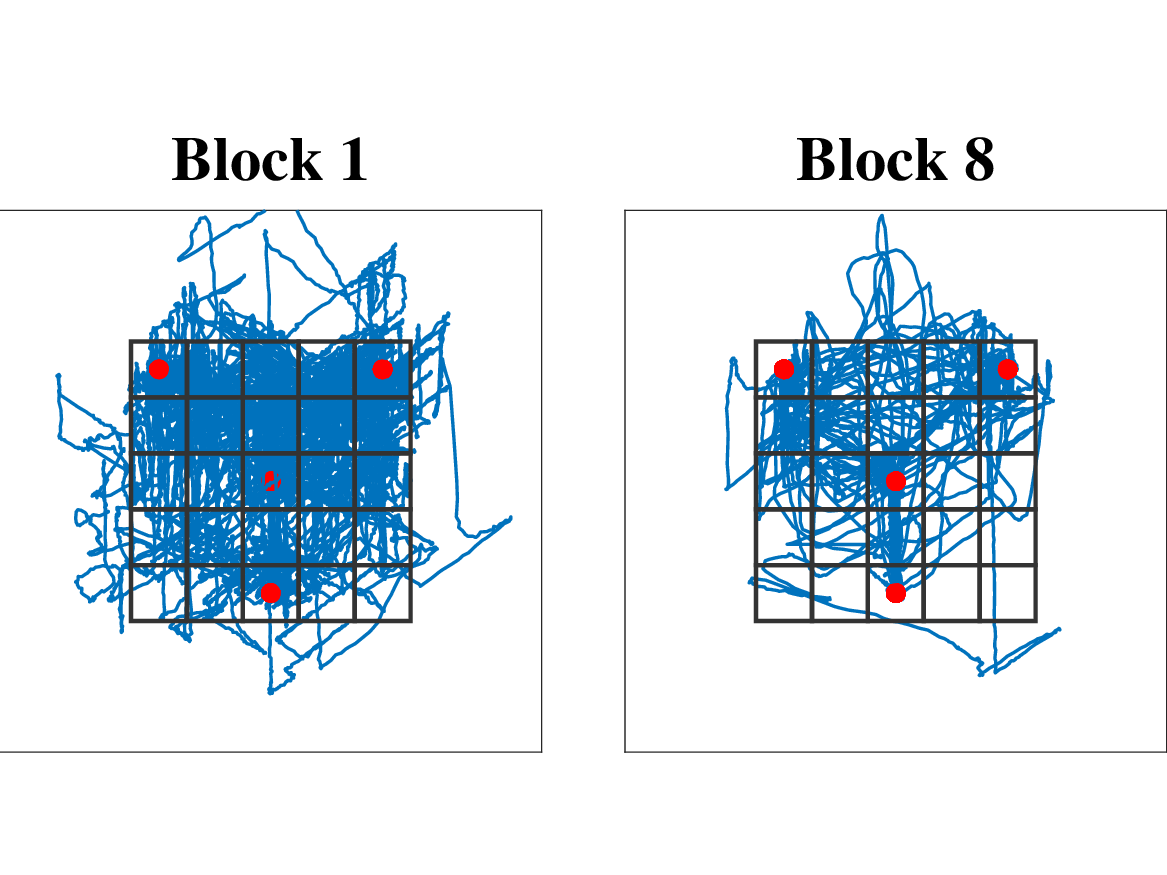}
    \end{subfigure}
    ~
    \centering
    \begin{subfigure}{0.3\linewidth}
	    \centering
        \caption{}
        \includegraphics[width=1\linewidth, height=1\linewidth, keepaspectratio]{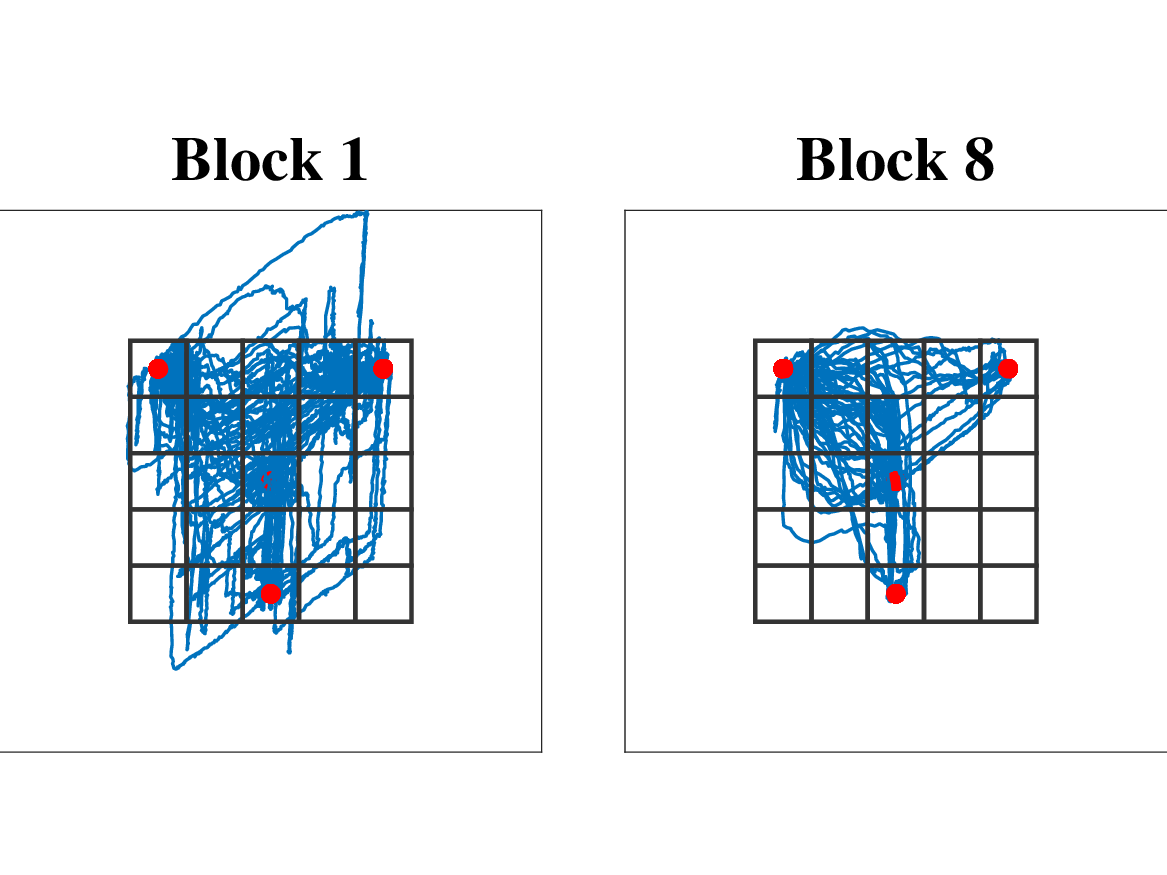}
    \end{subfigure}
    ~
    \centering
    \begin{subfigure}{0.3\linewidth}
	    \centering
        \caption{}
        \includegraphics[width=1\linewidth, height=1\linewidth, keepaspectratio]{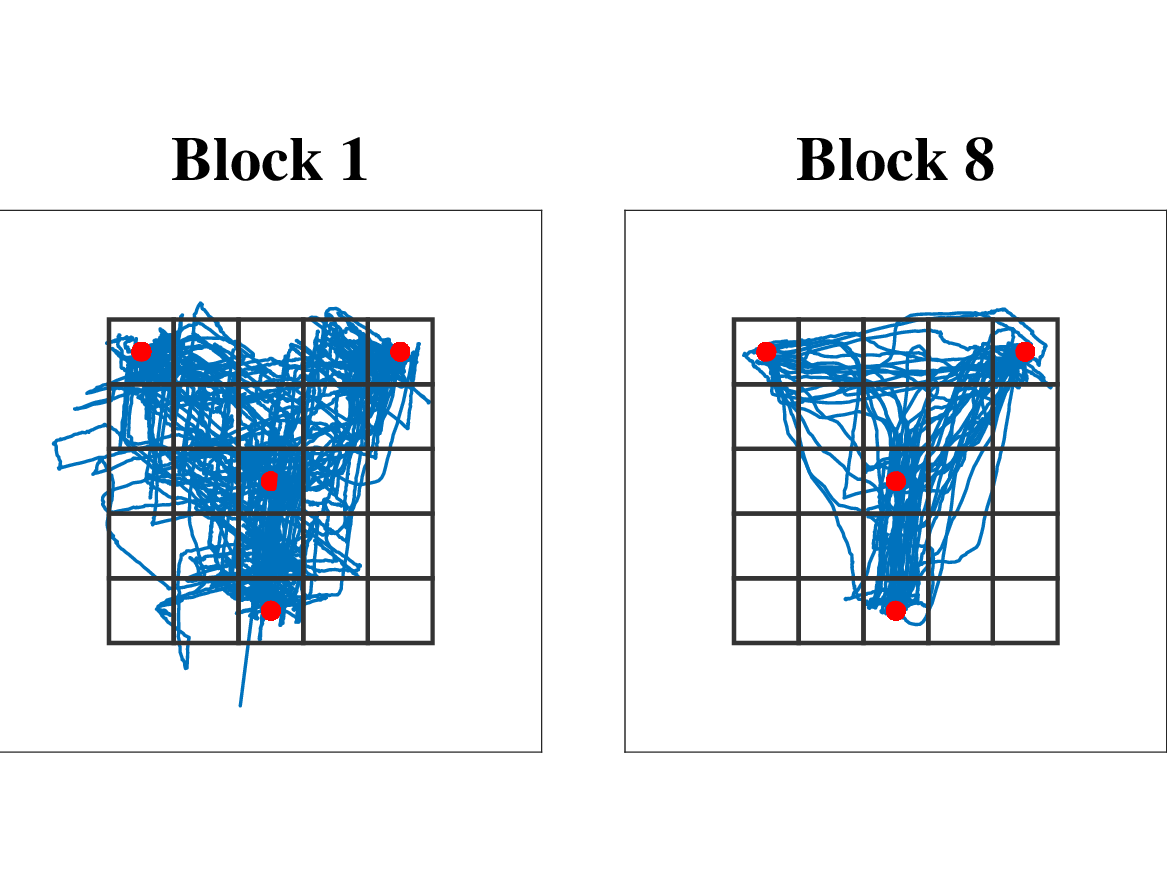}
    \end{subfigure}
    
    \caption{\textbf{Gameplay Trajectories:} Gameplay trajectories for the worst (top) and the best (bottom) participants in (a) control group, (b) manual group, and (c) SNMPC group show how the participants trained on SNMPC-based curriculum can achieve straighter trajectories towards the end of the training. The best and worst participants are selected based on the $\norm{\texttt{SoT}}_2$ values across the gameplay trials.}
    \label{fig:gameplay_trajs}
\end{figure*}

We show the gameplay trajectories of the best and the worst participants, based on $\norm{\texttt{SoT}}_2$ across the trials, in each experiment group. Fig. \ref{fig:gameplay_trajs} shows how SG participants achieve straighter trajectories between the different target-pairs by the end of the training session.
\section{Illustration of Performance Metrics }
\subsubsection{Straightness of Trajectory}
\begin{figure}[!h]
    \vspace{-0.1in}
    \centering
    \includegraphics[width=0.5\linewidth, trim=1cm 3.5cm 3cm 3cm, clip]{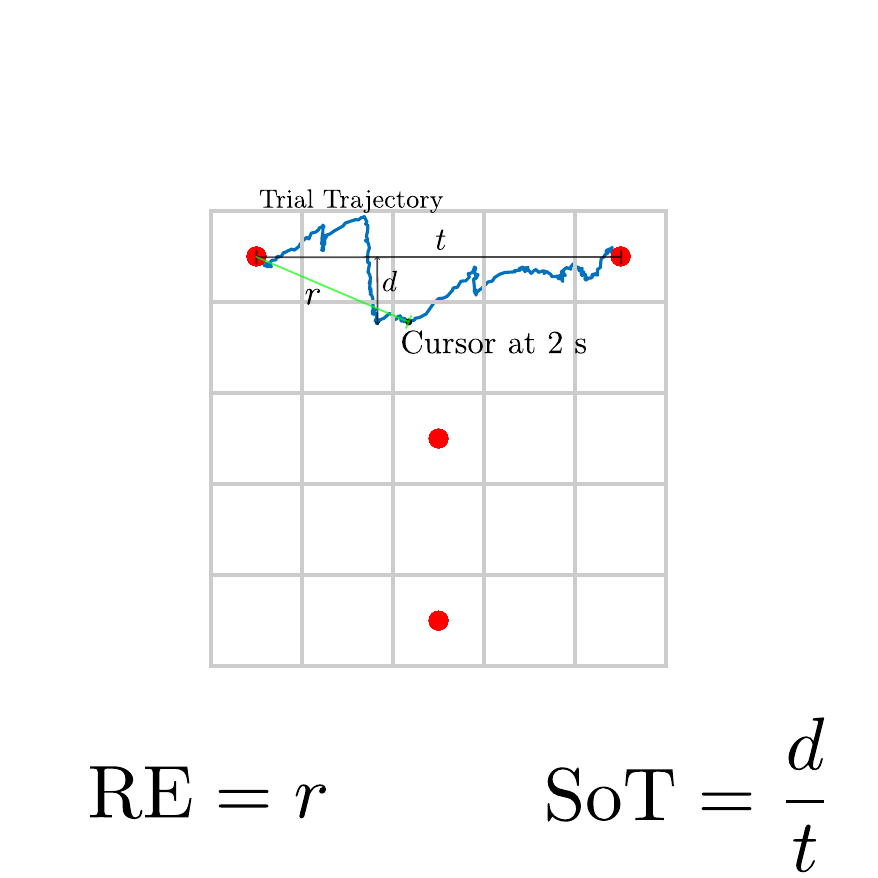}
    \caption{\textbf{Performance Metrics:} \texttt{SoT} is calculated as the ratio of maximum deviation from the straight line to the length of the straight line between the start and target points. While \texttt{RE} is computed as the Euclidean distance of the cursor from the target location at the end of movement or at $2~$s, whichever is earlier.}
    \label{fig:sot_illustration}
\end{figure}

Fig. \ref{fig:sot_illustration} shows how \texttt{SoT} is calculated as the maximum deviation of the cursor trajectory from the straight line joining the start and end points, normalized to the distance between the start and the end points.
\subsubsection{Reaching Error}
Fig. \ref{fig:sot_illustration} shows how \texttt{RE} is computed as the Euclidean distance of the cursor from the end point at the end of the trial or at $2~$s, whichever is earlier.

\section{Choice of Nonlinear Filter}
Our choice to employ a particle filter for nonlinear state estimation is motivated by the strong nonlinearities in the motor learning dynamics \eqref{appendix:HML_model}. Particularly, the equations \eqref{appendix:HML_model:W_hat} and \eqref{appendix:HML_model:u} contains products of multiple state variables $(\hat{W} \Phi \delta\bs q, \hat{W}\bs u)$ and outer products of states $(\delta \bs q \delta \bs q\tran, \hat{W}\tran \hat{W})$ that introduce strong nonlinearities. For such complex state interactions, the propagated sigma points for the UKF can end up in a configuration that is poorly represented by a single Gaussian. Thus, the resulting mean and covariance are off, making the filter estimates diverge. This issue is exacerbated in the EKF since it relies on linearization, which is an even poorer approximation of the dynamics. However, a PF's prediction step is simply a Monte Carlo simulation, and thus can capture the non-Gaussian distribution accurately. Moreover, maintaining a cloud of particles makes PF more robust to any divergences.
\begin{figure*}[h!]
    \centering
    \begin{subfigure}{0.34\linewidth}
	    \centering
        \caption{}
        \includegraphics[width=1\linewidth, height=1\linewidth, keepaspectratio]{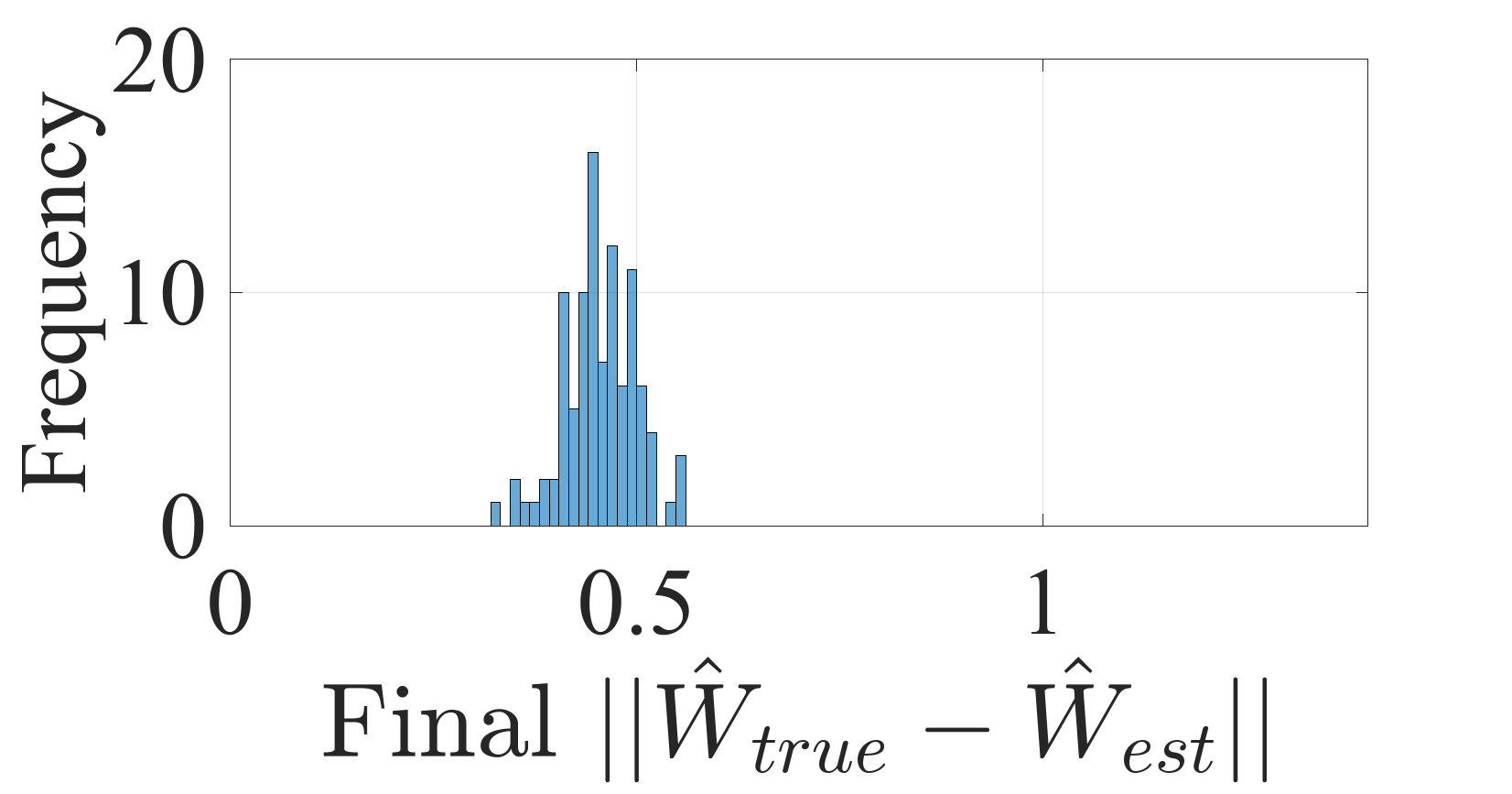}
    \end{subfigure}\hspace{-1.2em}%
    ~
    \centering
    \begin{subfigure}{0.34\linewidth}
	    \centering
        \caption{}
        \includegraphics[width=1\linewidth, height=1\linewidth, keepaspectratio]{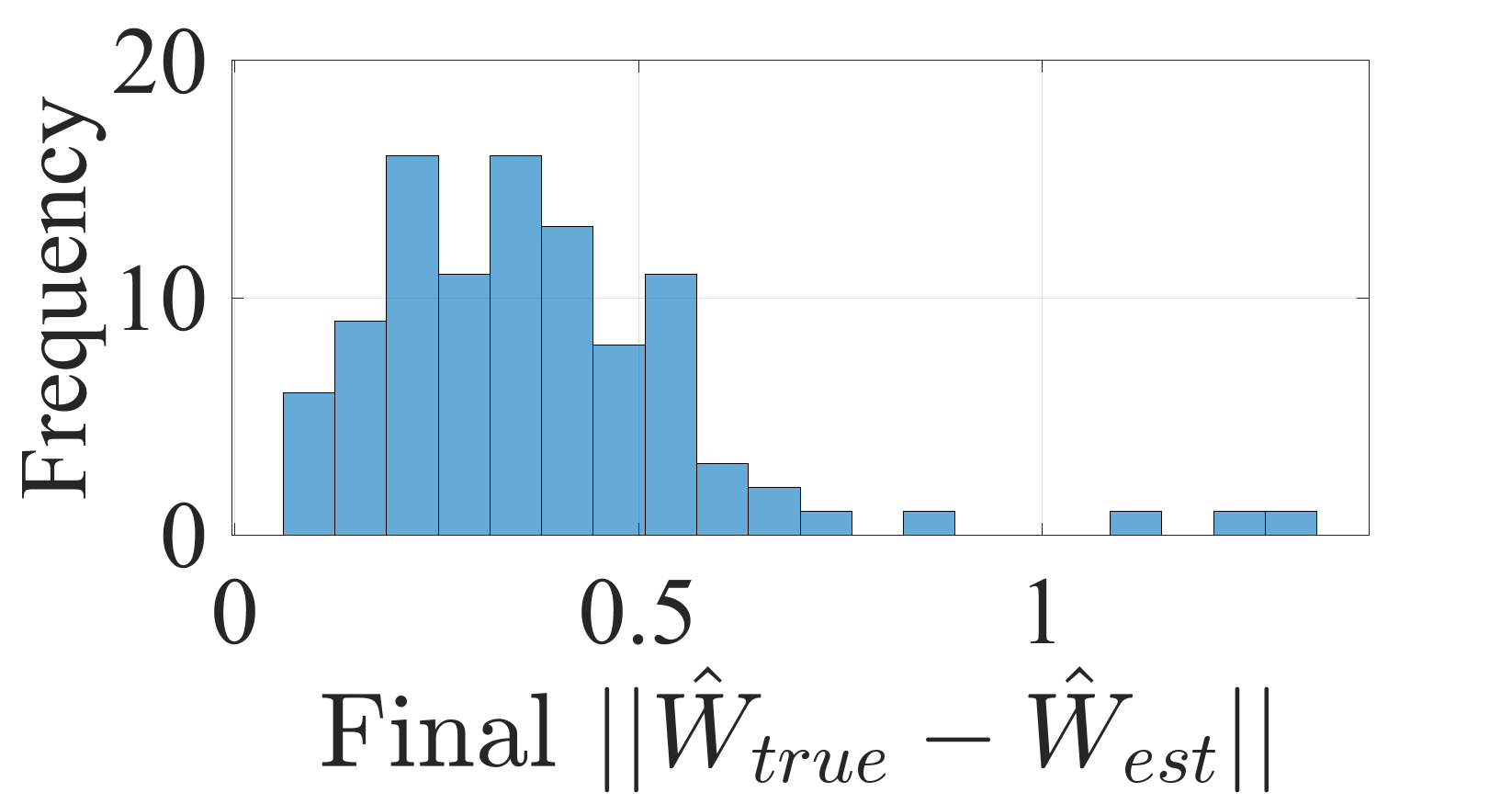}
    \end{subfigure}\hspace{-1.2em}%
    ~
    \centering
    \begin{subfigure}{0.34\linewidth}
	    \centering
        \caption{}
        \includegraphics[width=1\linewidth, height=1\linewidth, keepaspectratio]{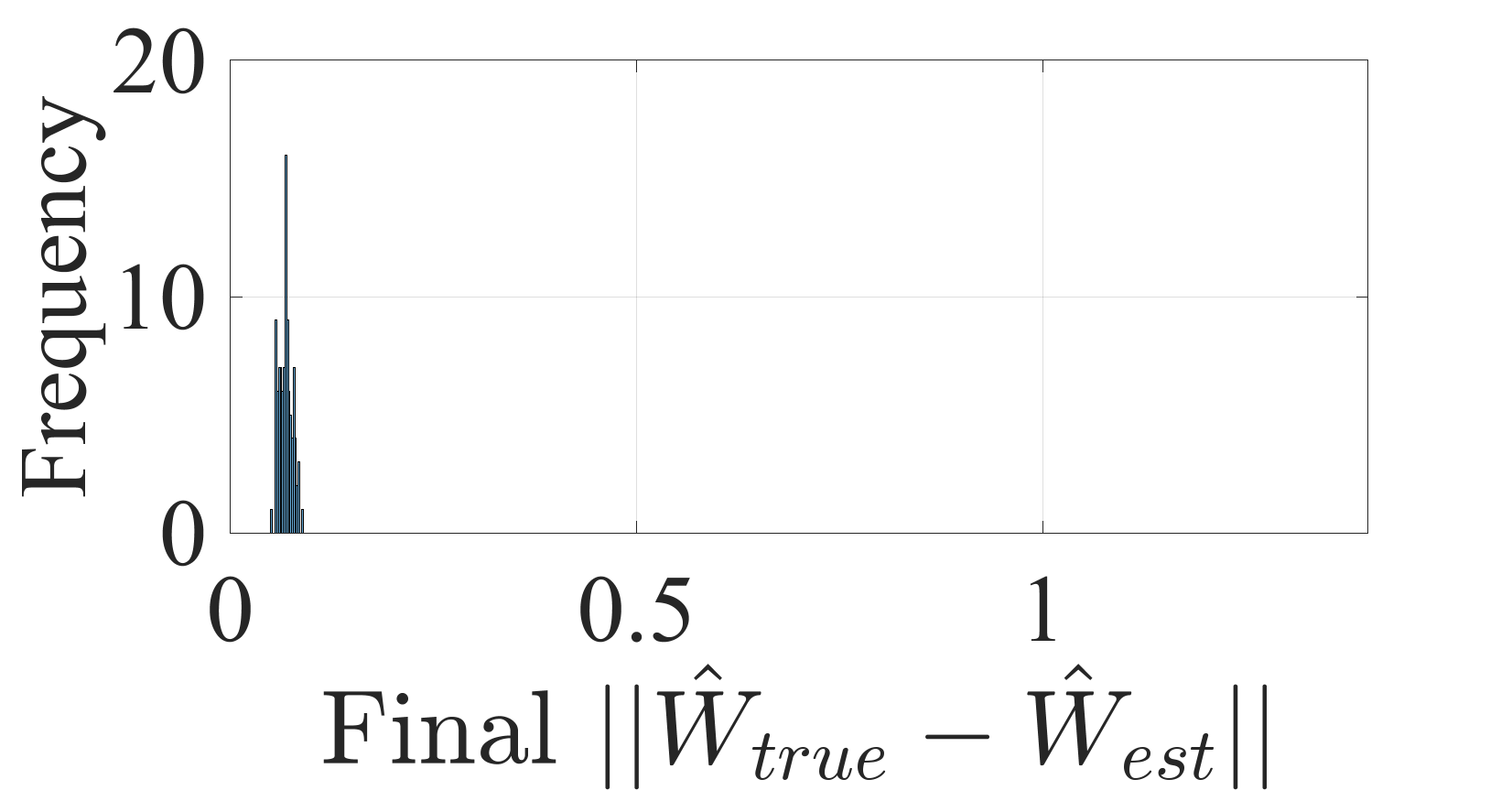}
    \end{subfigure}\hspace{-1.2em}%
    
    \caption{\textbf{Distribution of final state estimation errors for nonlinear filters} over $100$ Monte-Carlo simulations of one game trial shows that the (a) EKF has high precision but low accuracy, (b) UKF has low precision and low accuracy, and (c) PF has high precision and high accuracy in estimating the state $\hat{W}$.}
    \label{fig:filter_analysis}
\end{figure*}

To demonstrate the efficacy of PF over Gaussian filters for state estimation in our setup, we run a \emph{consistency analysis}. Specifically, we run all the filters for $100$ Monte-Carlo simulations, adding a random perturbation to the initial state each time, and then plot the distribution of the final state estimation errors over one trial of the simulated target capture game. Fig. \ref{fig:filter_analysis} shows that EKF has the largest final state estimation error consistently, whereas UKF has a relatively smaller state estimation error, but with a larger spread, evidence of low robustness. However, PF has the lowest and most consistent final state estimation error, showing how confident and accurate the state estimation is for our use case.
\end{document}